\documentclass[journal]{IEEEtran}
\ifCLASSINFOpdf
\else
\fi

\ifCLASSOPTIONcompsoc
    \usepackage[caption=false, font=normalsize, labelfont=sf, textfont=sf]{subfig}
\else
	\usepackage[caption=false, font=footnotesize]{subfig}
\fi

\usepackage{cite}
\usepackage[version=4]{mhchem}
\usepackage{graphicx}
\usepackage{amsmath}
\usepackage{amssymb}
\usepackage{algorithm}
\usepackage[noend]{algpseudocode}
\usepackage{tablefootnote}
\makeatletter
\def\BState{\State\hskip-\ALG@thistlm}
\makeatother

\begin{document}
\title{Adaptive Precision CNN Accelerator Using\\ Radix-X Parallel Connected Memristor Crossbars}

\author{Jaeheum~Lee,
        Jason~K.~Eshraghian,~\IEEEmembership{Member,~IEEE,}
        Kyoungrok~Cho,~\IEEEmembership{Senior~Member,~IEEE,}
        and~Kamran~Eshraghian
	\thanks{This work was partially supported by the National Research Foundation of Korea Grant funded by the Korean Government under Grant NRF-2017R1D1A1A09000613, in part by the DFAT Australia–Korea Foundation under Grant AKF00640, and in part by iDataMap Corporation, Australia.}
	\thanks{J. Lee and K. Cho are with the Department of Information and Communication Engineering, Chungbuk National University, Cheongju 362763, South Korea (e-mail: ljf9307@cbnu.ac.krl krcho@cbnu.ac.kr).}
	\thanks{J. K. Eshraghian is with the School of Electrical, Electronic and Computer Engineering, University of Western Australia, Perth, WA 6009, Australia (e-mail: jason.eshraghian@uwa.edu.au).}
	\thanks{K. Eshraghian is with iDataMap Corporation Pty. Ltd., Eastwood, SA 5063, Australia (e-mail: 	keshraghian@idatamap.com).}}%

\markboth{ }%
{Lee \MakeLowercase{\textit{et al.}}: Adaptive Precision CNN Accelerator Using Radix-X Parallel-Connected Memristor Crossbars}

\maketitle

\begin{abstract}
Neural processor development is reducing our reliance on remote server access to process deep learning operations in an increasingly edge-driven world. By employing in-memory processing, parallelization techniques, and algorithm-hardware co-design, memristor crossbar arrays are known to efficiently compute large scale matrix-vector multiplications. However, state-of-the-art implementations of negative weights require duplicative column wires, and high precision weights using single-bit memristors further distributes computations. These constraints dramatically increase chip area and resistive losses, which lead to increased power consumption and reduced accuracy. In this paper, we develop an adaptive precision method by varying the number of memristors at each crosspoint. We also present a weight mapping algorithm designed for implementation on our crossbar array. This novel algorithm-hardware solution is described as the radix-X Convolutional Neural Network Crossbar Array, and demonstrate how to efficiently represent negative weights using a single column line, rather than double the number of additional columns. Using both simulation and experimental results, we verify that our radix-5 CNN array achieves a validation accuracy of 90.5\% on the CIFAR-10 dataset, a 4.5\% improvement over binarized neural networks whilst simultaneously reducing crossbar area by 46\% over conventional arrays by removing the need for duplicate columns to represent signed weights. 
\end{abstract}


\IEEEpeerreviewmaketitle

\section{Introduction}
\IEEEPARstart{M}{achine} learning algorithms have become ubiquitous in the modern world, and are crucial in enabling computer systems which automatically update and improve with experience. This has opened up new frontiers in data analysis techniques. Deep learning refers to the use of a multi-layered neural network where the sequence of layers between the input and output perform feature identification at various hierarchies, as inspired by an approximation of the neuronal connections within the brain \cite{Yanagisawa2018, Khagi2018, Ushizima2016, Lawrence1997}. A popular deep learning algorithm for structured data is the convolutional neural network (CNN), which are well suited for object detection and vision-based processing, due to their high performance in feature recognition and object detection in images \cite{He2016}.

One of the challenges associated with machine learning stems from dimensionality issues, where algorithms with more features in higher dimensional spaces lead to difficulty in interpretability of the  network. When a learning algorithm does not work, the simplest path to success is often to feed the machine more data. This leads to scalability issues, where we have more data but lack the processing power to compute new inferences. An almost real-time prediction with sufficient accuracy is required for portable devices and edge sensors, using a constrained power budget to implement ambient-assisted technologies. 

\noindent This challenge was initially addressed by shifting computations over to graphical processing units (GPU), as GPU architectures consist of many small cores that parallelize the processing of data.  Calculations of similar form are carried out simultaneously, thus maximizing throughput of all threads which boosts performance while reducing the bottleneck when paired with a CPU. However, when dealing with algorithms that must call a significant number of parameters from memory, (e.g., 138 million parameters in the VGG-16 CNN \cite{Simonyan2014}), these parameters must be accessed from and stored in data memory via a shared bus with restrictive data transfer rates. This issue is referred to as the von Neumann bottleneck.

\noindent More recently, application specific Neural Processing Units (NPUs) were deployed in mobile devices for real time operation without the need for server connections to perform deep learning operations \cite{Kiningham2016, Hong2018, Chen2016, Wang2018}. NPUs are optimized for power and area efficiency for matrix-vector multiplication (MVM) without the need for `cloud-based' processing. However, this approach still relies on conventional CMOS technology where process scaling is bound to performance degradation (retention, cycling and reliability), and memory and processing are physically delocalized. This has given rise to the exploration of beyond-CMOS architectures for artificial neural network (ANN) and CNN applications. 

Researchers have offered a variety of hardware solutions that implement memristors into neuromorphic processors \cite{Zhang2018, Burr2017, Yang2017, Li2018, Liu2017, Zhao2017, Eshraghian2018, Hu2014}. The memristor is a two-terminal nanoscale device which serves as non-volatile memory and also doubles as a resistor. That is, memory and computation based on the linear form of Ohm's Law exist within the same device. Memristors are scaled into a dense crossbar structure for an area efficient means to parallelize  multiply-and-accumulate (MAC) functions, where high-speed computation is achieved through the column-wise parallelism of arrays. However, problems such as memory leakage, variability and device sensitivity make it challenging to reliably store multi-bit and analog data \cite{Eshraghian2016, Ni2016, Stathopoulos2017, Tang2017}. The work in \cite{Li20182} demonstrates the storage of over 64 conductance states per memristor, though the difference between simulated and experimental efficiency is an order of magnitude of $10^2$ TOPS/w, speculated to be a result of the slow write times needed to ensure precise conductance control and noise mitigation.

To combat the limitation of multi-bit and analog state memristors, hardware implementations using two states ($R_{ON}$, $R_{OFF}$) of a memristive binarized neural network (BNN) \cite{Courbariaux2016, Krestinskaya2018} model have been proposed. Where weights are limited to single-bit resistances, lower precision results in decreased classification accuracy. Other methods to achieve multi-bit weights are through binarized encoding schemes with column-wise distribution, or via frequency modulation by encoding weight information in the time-domain of the driving voltage \cite{Eshraghian2019}. In all cases, either chip area or timing is compromised due to additional columns and the need for more complex CMOS driving circuitry. The representation of negative weights with positive conductances requires double the number of columns, with outputs passed through a differential amplifier \cite{Li20182}.

In this paper, we propose a novel solution derived from nanoelectronics to overcome the above limitations of conventional crossbar architectures. This is done by introducing parallel-connected memristors at each crosspoint junction on a crossbar, by either splitting larger memristors and insulating the smaller counterparts from one another, or laying out multiple masks per crosspoint. This means we are able to process radix-X weights (i.e., higher bit precision at each junction), and formalize a hardware mapping approach that significantly reduces circuit area utilization by representing both negative and positive weights without the need to distribute computations across column wires. Furthermore, this approach significantly reduces exposure to line losses.

The main contributions of this paper are:

\begin{enumerate}
	\item \textbf{Radix-X CNN:} here we introduce a CNN implementation using radix-X weights, where the weights and activation values are mapped to the range of the radix numeral system, or `X'. We develop a straightforward algorithm based on regularization, and provide both pseudo-code and our python implementation. We test the accuracy of our radix-X CNN by training it on the CIFAR-10 dataset, and comparing it with several prominent models. Intuitively, this can be thought of as targeting the algorithmic component in algorithm-hardware co-design methodology.
	
	\item \textbf{Parallel-connected memristors at each crossbar junction for radix-X weight representation:} hardware implementation of radix-X CNN. We show improved stability, reliability and decreased area consumption by using our proposed parallel-connected memristor architecture for storage of radix-X CNN weights. This focuses on the hardware aspect of the co-design methodology.
	
	\item \textbf{Negative weights representation:} implementation of negative weights is a significant overhead in crossbar arrays. Conventional methods use twice the area of crossbars to address this problem. Here, we demonstrate how our radix-X CNN significantly reduces the circuit area by using a single crossbar reference column for both negative and positive weight representation, rather than doubling the number of column wires.
\end{enumerate}

\noindent The above contributions are quantified by showing how our proposed radix-X CNN hardware achieves a validation accuracy of 90.5\% on the CIFAR-10 dataset when $X=5$, and a 4.5\% improvement on conventional low precision weights (namely, BNNs). Importantly, we reduce chip area by 46\% over conventional state-of-the-art arrays by condensing the number of required column wires to represent negative weights down to a single reference line.

This paper is organized as follows: section II introduces the concepts that drive the technology of the radix-X CNN approach in a memristor crossbar. Section III describes our radix-X CNN learning algorithm with pseudo-code provided, and section IV demonstrates how it is implemented using a parallel-connected memristive crossbar array for representation of radix-X weights, and proposes a solution for negative and multi-bit weight representation. Section V shows our simulation results by running a classification example on the CIFAR-10 dataset, and section VI presents the nanofabrication techniques employed in the development of our crossbar array, with accompanied experimental results of a simple convolutional kernel with a Sobel filter containing both positive and negative elements applied to an input image. Section VII provides a discussion of some of the design trade-offs of the hardware implemented radix-5 CNN, with concluding remarks given in section VIII.

\section{Background}

\subsection{Resistive Switching in Memristors}
The reconfigurability of conductance in a memristor is leveraged in neuromorphic computing to represent updatable weight values. Resistive switching has been demonstrated in metal-oxide devices, with \ce{Ta2O5} \cite{Lee2011, Torrezan2011}, \ce{HfO2} \cite{Murdoch2016} and \ce{TiO2} \cite{Strukov2008, Yang2008} being among the most recognized. Under the influence of an applied electric field, a conductive filament made up of oxygen vacancies can be formed which creates a pathway for electrons to flow through \cite{Kwon2010}. The formation of the filament corresponds to a low resistance, and the rupture of the filament breaks the conductive pathway resulting in a high resistance.

\begin{figure}
	\centering
\subfloat[]{\raisebox{8mm}{\includegraphics[width=0.4\linewidth]{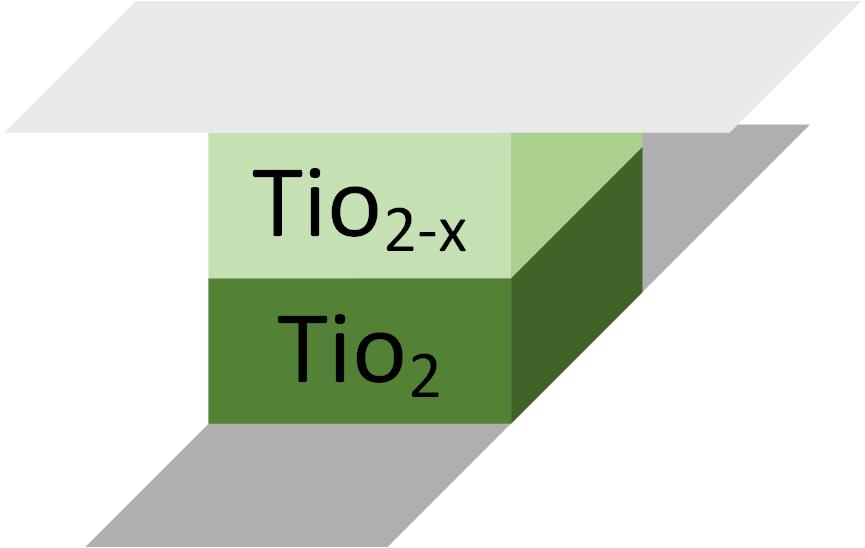}}\label{Fig1a}}
\hfil
\subfloat[]{{\includegraphics[width=0.6\linewidth]{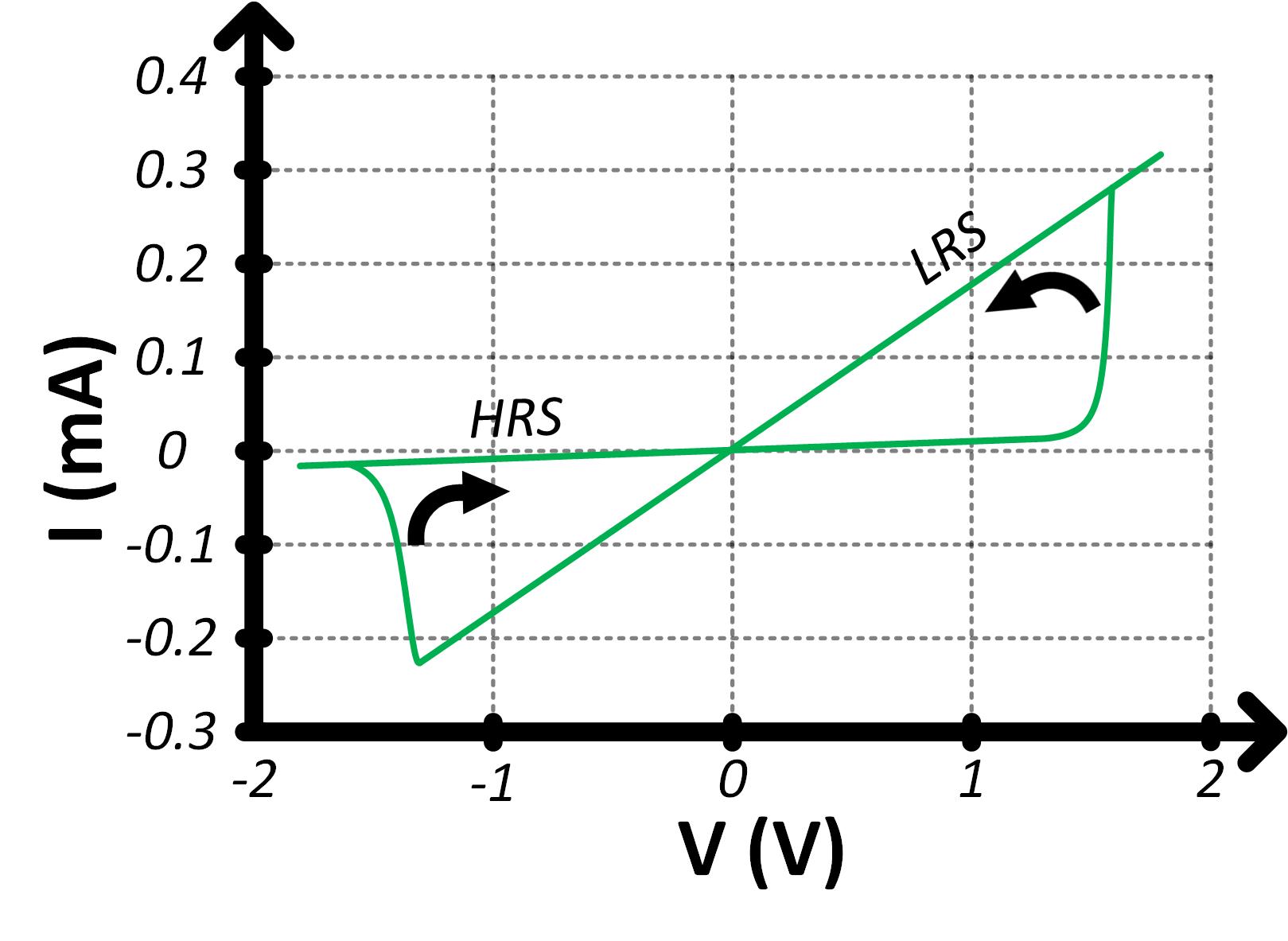}}}
\caption{Memristor characterization (a) physical representation depicting \ce{TiO2} and \ce{TiO_{2-x}}. (b) V-I characteristics of the model illustrating the LRS and HRS. Resistance ratio of this memristor is 100 when a sinusoidal voltage with frequency = 1 kHz is applied.}
\label{Fig1}
\end{figure}

Under a forward bias, the memristor switches to a low resistance state (LRS). When a reversal of the bias is applied, it switches to a high resistance state (HRS). Fig. \ref{Fig1}(a) illustrates the physical structure of a memristor formed by \ce{TiO2} and oxygen deficient \ce{TiO_{2-x}} layers sandwiched between two metal electrodes. Fig. \ref{Fig1}(b) illustrates the resultant V-I curve under a sinusoidal driving voltage, causing the device to switch between two resistance states. 

To achieve analog or multi-bit states, the width of the filament must be precisely modulated, which is challenging in practice. It often requires the use of lower write voltages applied across longer durations, which super-exponentially increase the time of write cycles \cite{Fuller2019}. Therefore, many realizations of crossbar arrays employ conservative design techniques and treat metal-oxide memory cells as single-bit storage \cite{Eshraghian2017}. Multi-bit weights are often implemented using multiple memristors, distributed across multiple column wires. 

\subsection{Convolutional Neural Networks}

\begin{figure}
\centering
\includegraphics[width=3.45in]{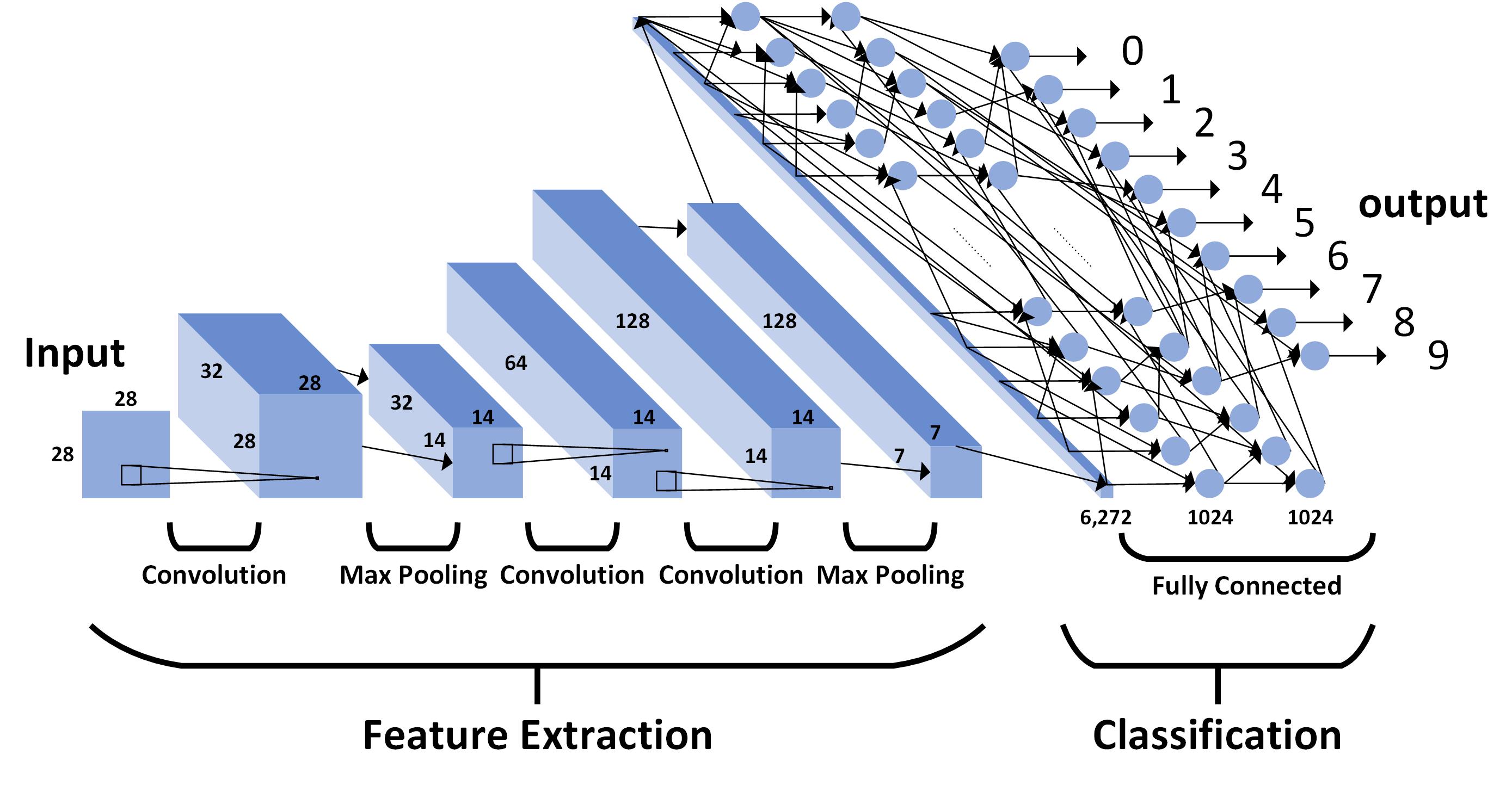}
\caption{Generalized CNN model with parameters labeled including number of layers, kernel size, and channel depth. Various CNN models can be created by altering these parameters and layer structures.}
\label{Fig2}
\end{figure}

A generic structure of a CNN is depicted in Fig.~\ref{Fig2} \cite{LeCun1989, Fukushima1980}. Its high performance in image classification is enabled by retaining some spatial dependencies (i.e., taking consideration of the location of pixels relative to neighboring pixels). This is achieved by treating the image as a matrix rather than vectorizing it in a fully-connected neural network. As higher-level features are extracted, the channel depth increases, which results in a much larger number of MVMs (computational equivalent of a MAC operation) for a given number of inputs. 

\subsection{Neural Network Using Memristor Crossbar Arrays}
The key to memristor crossbar arrays being capable of neural network acceleration is that MVMs are the dominant process in CNNs. By parallelizing a large number of MACs across column wires using weights that have been stored in the form of conductance values, we are able to optimize the hardware mapping of neural network architectures.  

\begin{figure}
	\centering
\subfloat[]{{\includegraphics[width=0.35\linewidth]{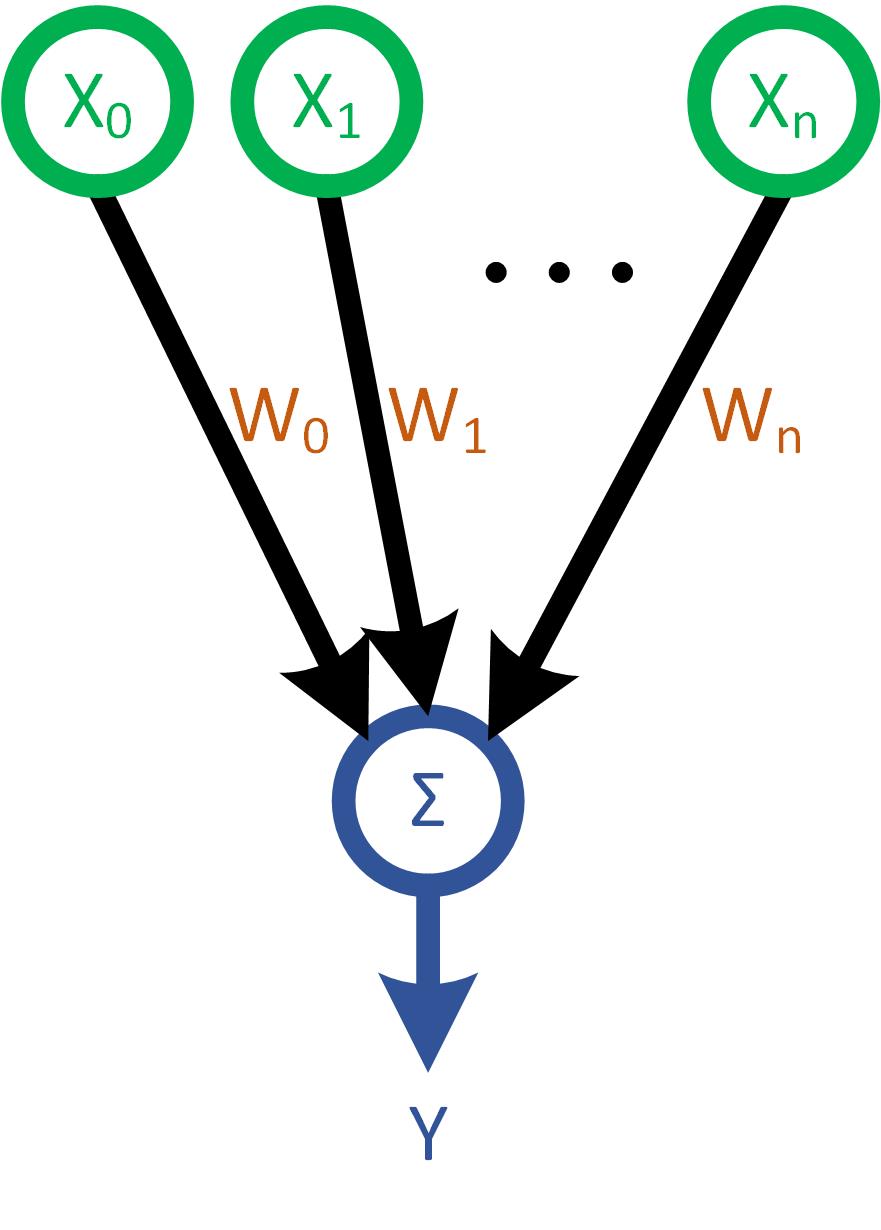}}}
\hfil
\subfloat[]{{\includegraphics[width=0.28\linewidth]{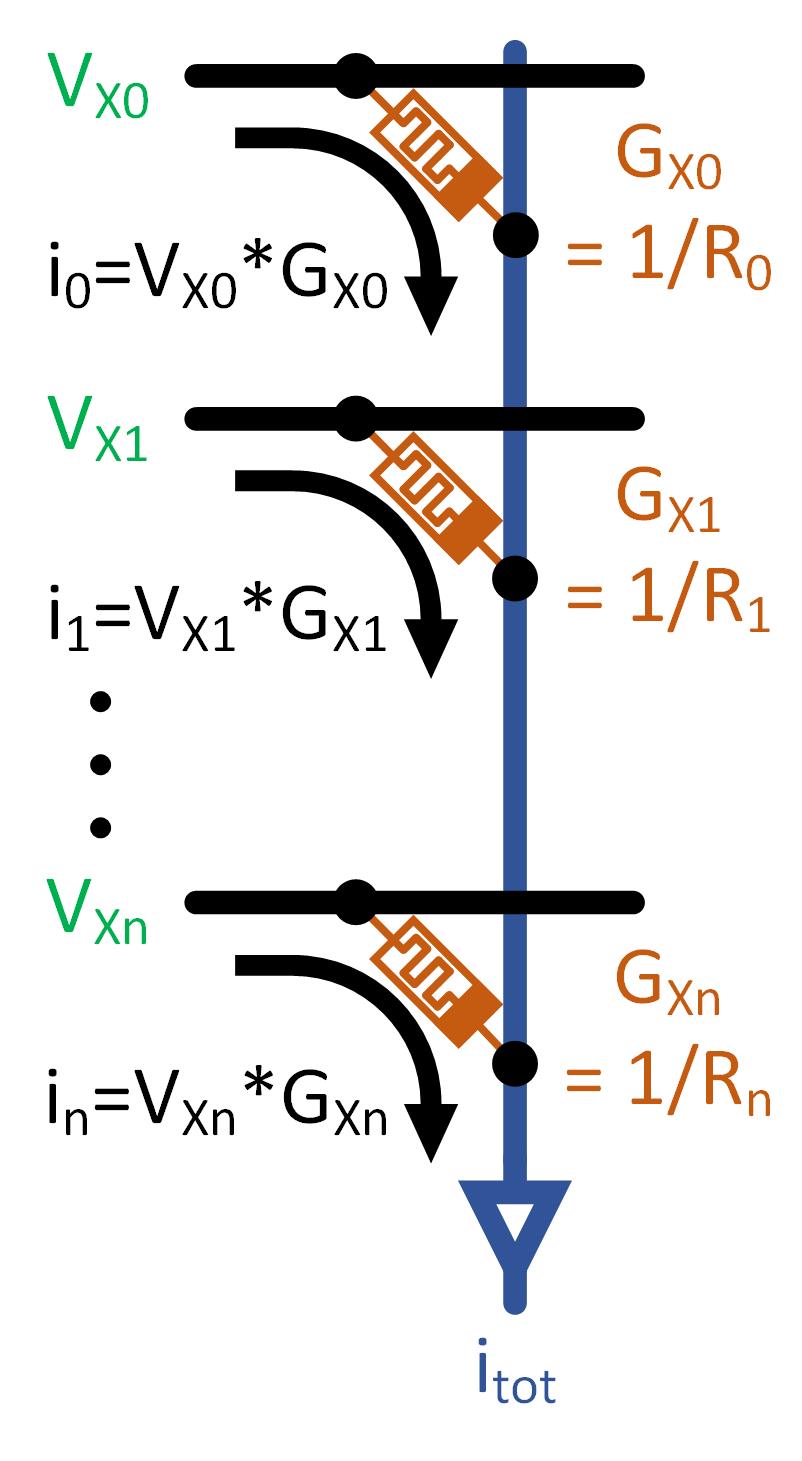}}}
\caption{The artificial neuron (a) architecture (b) mapping to a single column within a memristor crossbar architecture.}
\label{Fig3}
\end{figure}

Figs.~\ref{Fig3}(a) and (b) depict the mapping of the neuron model to a circuit. The inputs of the neural network $X_0$ to $X_n$ are linearly mapped to the input voltages $V_{x0}$ to $V_{xn}$ of the crossbar, and the weights $w_0$ to $w_n$ are linearly mapped to the conductances $G_{x0}$ to $G_{xn}$ of the memristor. By using the virtual ground of an inverting amplifier to hold each column wire as the reference node (detailed in section IV), the current drawn by each memristor can be calculated using Ohm's Law, and then summed along the column wire in accordance to Kirchhoff's Current Law. Equations (\ref{Eq1}) and (\ref{Eq2}) mathematically describe this process:

\begin{equation} \label{Eq1}
Y = \sum_{i=0}^{n} X_i * w_i,
\end{equation}

\begin{equation} \label{Eq2}
i_{tot} = \sum_{i=0}^{n} V_i * G_i,
\end{equation}

\noindent where $Y$ is the pre-activation output of the artificial neuron, $n$ corresponds to the number of inputs to the neuron, and $i_{tot}$ is the total current  through a column wire. A vectorized implementation of Fig.~\ref{Fig3}(b) is defined by (\ref{Eq3}).  When the number of columns is increased to a value $m$ in an array, (\ref{Eq2}) can be extended to MVM in (\ref{Eq4}):

\begin{equation} \label{Eq3}
i_{tot} = 
\begin{bmatrix} 
V_0 & V_1 & ... & V_n 
\end{bmatrix} *
\begin{bmatrix} 
G_0 \\ G_1 \\ \vdots \\ G_n
\end{bmatrix}
\end{equation}

\begin{equation} \label{Eq4}
\begin{bmatrix} 
i_0 & i_1 & ... & i_m
\end{bmatrix} = 
\begin{bmatrix} 
V_0 \\ V_1 \\ \vdots \\ V_n 
\end{bmatrix}^T 
\begin{bmatrix} 
G_{0,0} & G_{0,1} & ... & G_{0,m} \\
G_{1,0} & G_{1,1} & ... & G_{1,m} \\ 
\vdots & \vdots & \ddots & \vdots \\ 
G_{n,0} & G_{m,1} & ... & G_{n,m}
\end{bmatrix}
\end{equation}

\noindent The conductance weights in a single column in the crossbar array correspond to a single channel in a CNN kernel. One can implement deep-channel kernels in parallel by distributing these across column wires. The voltage corresponding to the image data is applied at the input terminals of the crossbar (i.e., at the row wires), where the convolution operation is performed. 

\section{Radix-X CNN Algorithm}
The conventional methods of working around single-bit weight restrictions in memristor crossbar arrays are either algorithmically by using BNNs, or via hardware distribution of computations via binarized encoding across columns. As mentioned, the former compromises accuracy and the latter expands chip area and power consumption. In the past, BNNs have been implemented either at the weight level, at the activation level (akin to the classic perceptron \cite{Rosenblatt1958}), and both in unison. The work in \cite{Courbariaux2016, Courbariaux2015, Rastegari2016} implements a binarized activation that can adopt both positive and negative values:

\begin{equation} \label{Eq5}
  x_b= \text{sgn}(x) := \begin{cases}
    +1, & \text{if $x\geq 0$}\\
    -1, & \text{otherwise}.
  \end{cases}
\end{equation}

\noindent Although this bounding approach is convenient for digitized implementation, there is a degradation of inference accuracy as a result of high precision compression \cite{Zhu2016}. This may be counteracted by using more learning parameters with an increased number of training epochs, but this offsets the advantages of parallelization. In light of these limitations, we propose a novel approach based on a radix-X weight representation, and present our method for algorithm and hardware co-design to realize it on a memristor crossbar array. 

\noindent The radix of a digital numeral system refers to the number of unique digits used to represent values in a positional numeral system, including the digit zero. If X is the radix of a numeral system, then in context of a neural network, radix-X refers to the complete set of values that are assignable as a weight and activation value. For example, where X=5, then in a radix-5 CNN the weights and activations can take on any one of 5 values. We present an algorithm that normalizes a high-precision pre-trained weight matrix into a radix-X weight matrix that can be any one of the values in the set \{-2, -1, 0, 1, 2\}. By employing the ReLU activation function, we ensure the outputs can also be represented within the limits of the radix-X numeral system as one of the values in the set \{0, 1, 2, 3, 4\}. In the most generalized case for radix-X, we propose that the weights must first be normalized according to the pseudo-code:

\begin{algorithm}
\caption{Convert pre-trained weights into radix-X}\label{euclid}
\begin{algorithmic}[1]
\Function{NormalizedTensor}{\textit{x}, \textit{weights}}:
\Comment Return normalized weights given input of radix-x and pre-trained weights
\State $\textit{range} \gets \text{max of}~\textit{weights} - \text{min of}~\textit{weights}$
\State $normalized \gets (\textit{weights} - \text{min of}~\textit{weights}) / \textit{range}$
\State $\textbf{return}~\textit{normalized}$
\EndFunction
\Function{QuantizedTensor}{\textit{weights}}:
\Comment Return quantized weights given input of pre-trained weights
\For{\textit{element} in \textit{weights}}:
\If{\textit{element} $<$ 0} round \textit{element} up
\Else~round \textit{element} down
\EndIf
\EndFor
\State $\textbf{return}~\textit{quantized}$
\EndFunction
\Comment In main function
\State \textbf{return} \textsc{QuantizedTensor}(\textsc{NormalizedTensor}(\textit{x, weights}))
\end{algorithmic}
\end{algorithm}

\noindent An integer input $x$ is the radix of a numeral system, and a matrix or tensor input of pre-trained weights, plainly denoted $weights$, are both passed into the function \textsc{NormalizedTensor}, which returns a normalized set of weights where the minimum element is $-X/2$ and the maximum element is $+X/2$. The output is called as the argument of \textsc{QuantizedTensor} which quantizes all floating point decimals into integers. For accessibility, we have provided a link to the GitHub repository containing our Python 3 implementation of the above pseudo-code \cite{EshraghianGit}. The Python code also includes the radix-X activation function.

\noindent Where $X=5$, the mathematical equivalent for a radix-5 CNN of the above algorithm is:

\begin{equation} \label{Eq6}
range = w_{max} - w_{min}
\end{equation}

\begin{equation} \label{Eq7}
w_{r-5}(w) = \begin{cases}
    -2, & \text{if $\frac{range}{5}* 0 \leq w - w_{min} < \frac{range}{5}*1$}\\
    -1, & \text{if $\frac{range}{5}* 1 \leq w - w_{min} < \frac{range}{5}*2$}\\
    0, & \text{if $\frac{range}{5}* 2 \leq w - w_{min} < \frac{range}{5}*3$}\\
    +1, & \text{if $\frac{range}{5}* 3 \leq w - w_{min} < \frac{range}{5}*4$}\\
    +2, & \text{if $\frac{range}{5}* 4 \leq w - w_{min} < \frac{range}{5}*5$}\\
  \end{cases}
\end{equation}

\begin{equation} \label{Eq8}
\begin{aligned}
pixel_{r-5}&=\text{ReLU}_{r-5}(pixel)   =  \\ &\begin{cases}
    0, & \text{if $pixel \leq \frac{pixel_{max}}{4}*0$}\\
    +1, & \text{if $\frac{pixel_{max}}{4}* 0 \leq pixel < \frac{pixel_{max}}{4}*1$}\\
   	+2, & \text{if $\frac{pixel_{max}}{4}* 1 \leq pixel < \frac{pixel_{max}}{4}*2$}\\
    +3, & \text{if $\frac{pixel_{max}}{4}* 2 \leq pixel < \frac{pixel_{max}}{4}*3$}\\
    +4, & \text{if $\frac{pixel_{max}}{4}* 3 \leq pixel < \frac{pixel_{max}}{4}*4$}\\
    \end{cases}
    \end{aligned}
\end{equation}

\begin{figure}
	\centering
\subfloat[]{{\includegraphics[width=0.8\linewidth]{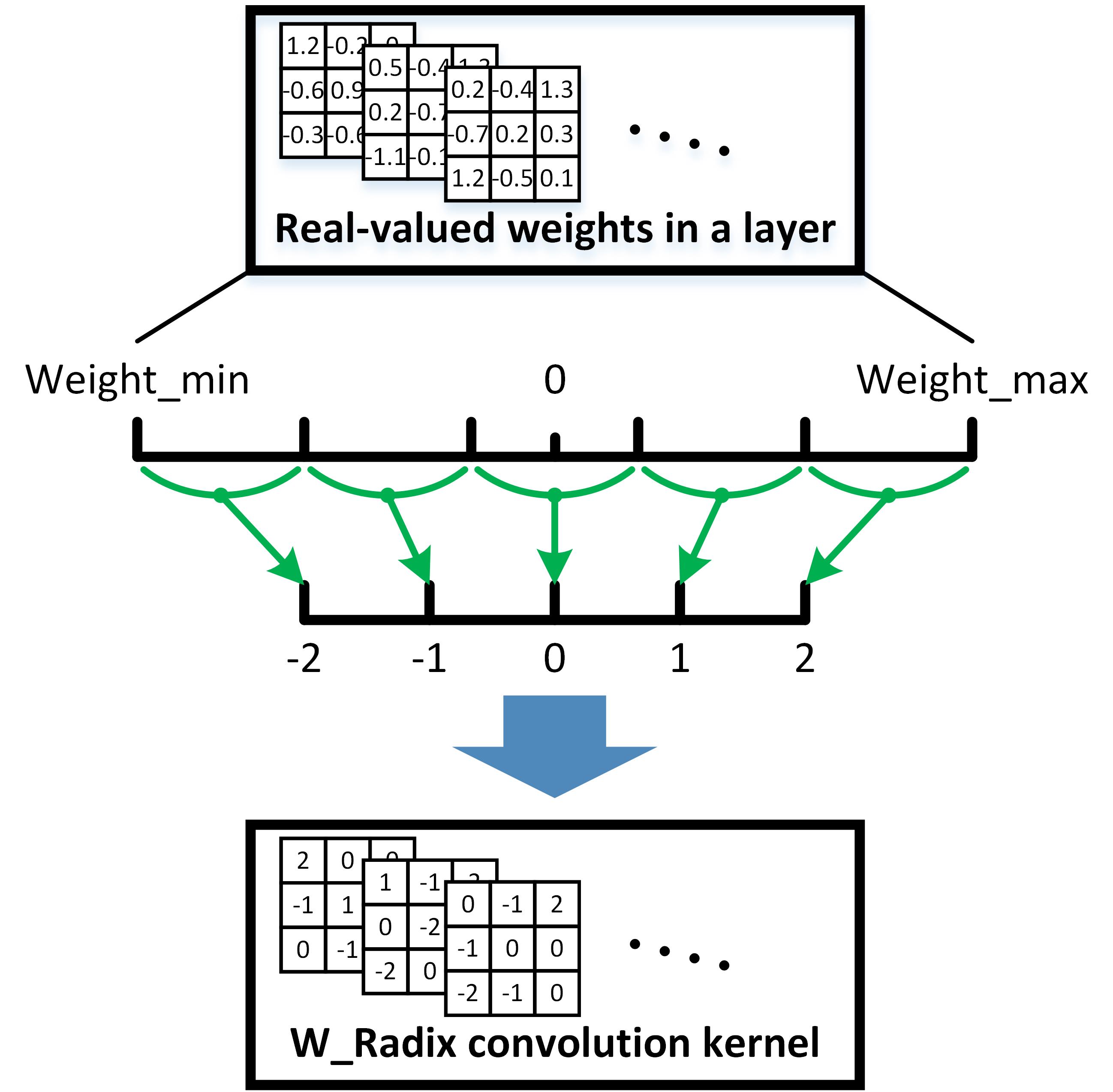}}}
\hfil
\subfloat[]{{\includegraphics[width=0.8\linewidth]{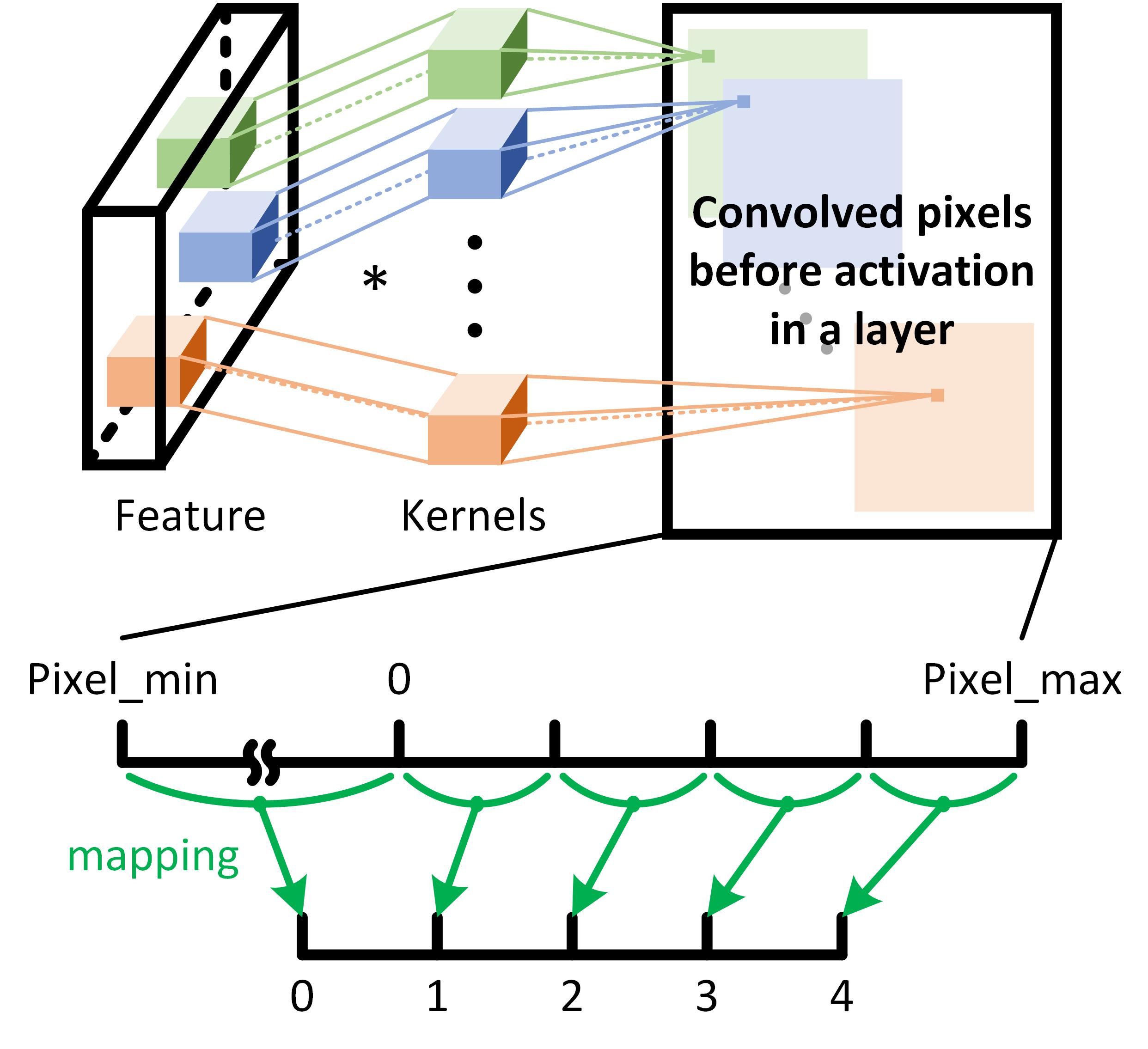}}}
\caption{Kernel mapping platform (a) process of converting a weight used in convolution to five limited numbers. The kernel values in a layer are normalized then quantized to \{-2, -1, 0, 1, 2\}. (b) Conversion is based on a bounded ReLU activation, where all values less than 0 are mapped to zero.}
\label{Fig4}
\end{figure}

\begin{figure}
	\centering
\subfloat[]{{\includegraphics[width=0.5\linewidth]{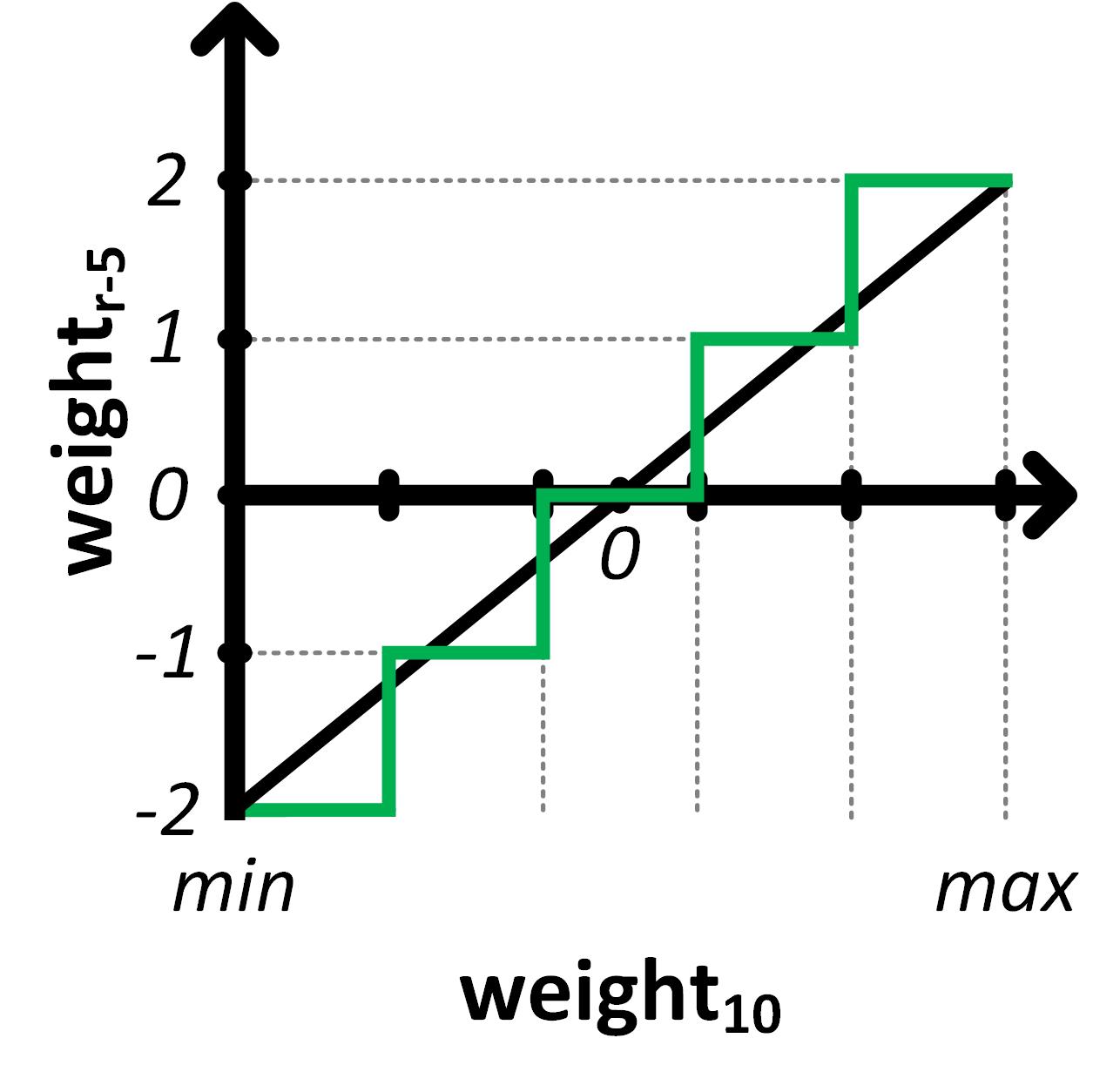}}}
\hfil
\subfloat[]{{\includegraphics[width=0.5\linewidth]{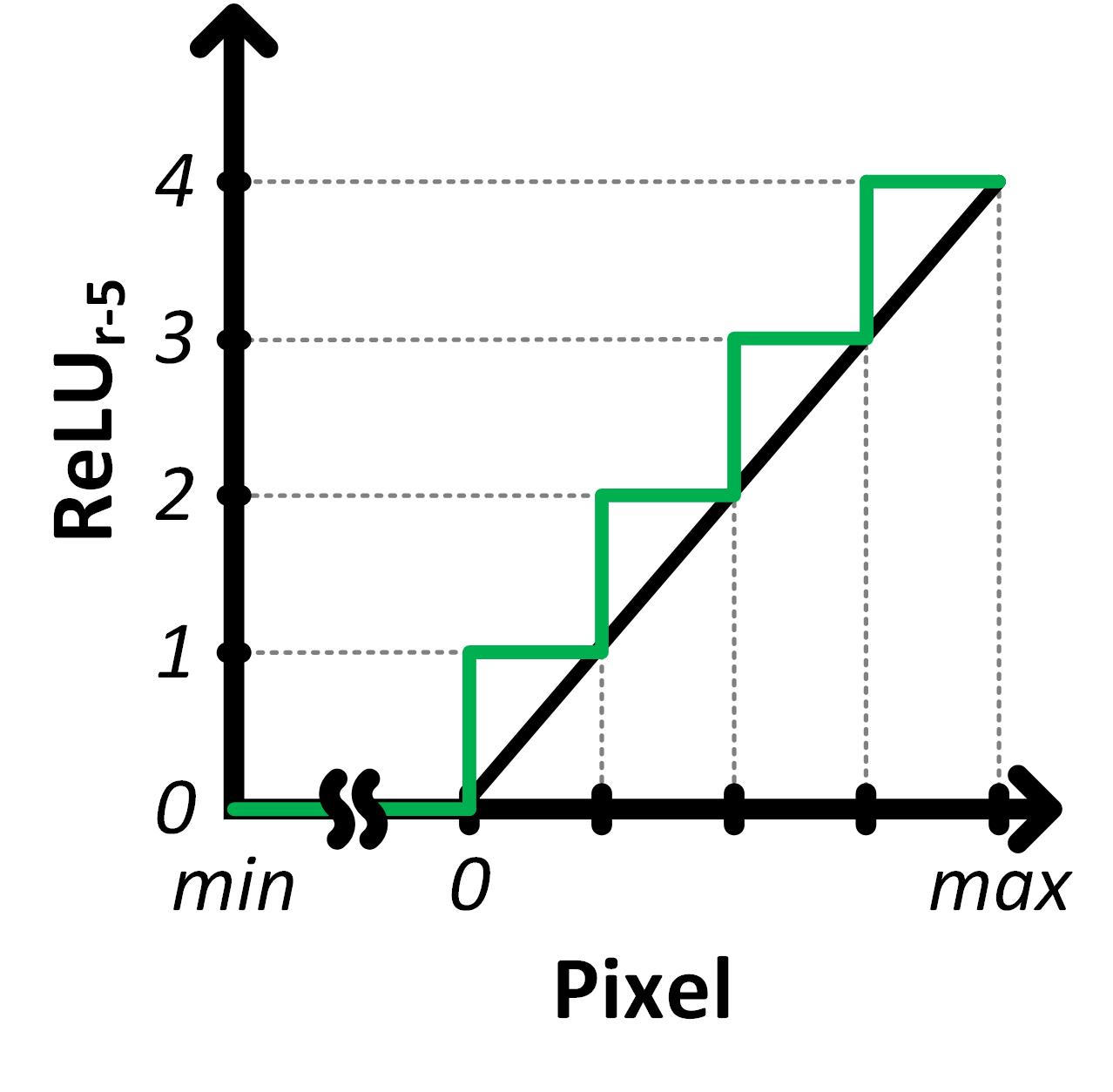}}}
\caption{Conversion plot of (a) weights and (b) activations.}
\label{Fig5}
\end{figure}

\noindent where $range$ refers to the range of weights prior to normalization, and $w_{max}$ and $w_{min}$ are the maximum and minimum weights respectively. In (\ref{Eq7}), $w_{r-5}$ is the equivalent set of weights in radix-5, and $w$ refers to the weights in the base 10 system. In (\ref{Eq8}), `pixel' is the input data convolved with a kernel before activation, and when passed through the $\text{ReLU}_{r-5}$ activation, gives an output of $pixel_{r-5}$, which is bound to one of five integer values. Figs.~\ref{Fig4} and \ref{Fig5} illustrate the process.

\begin{figure}
\centering
\includegraphics[width=3.45in]{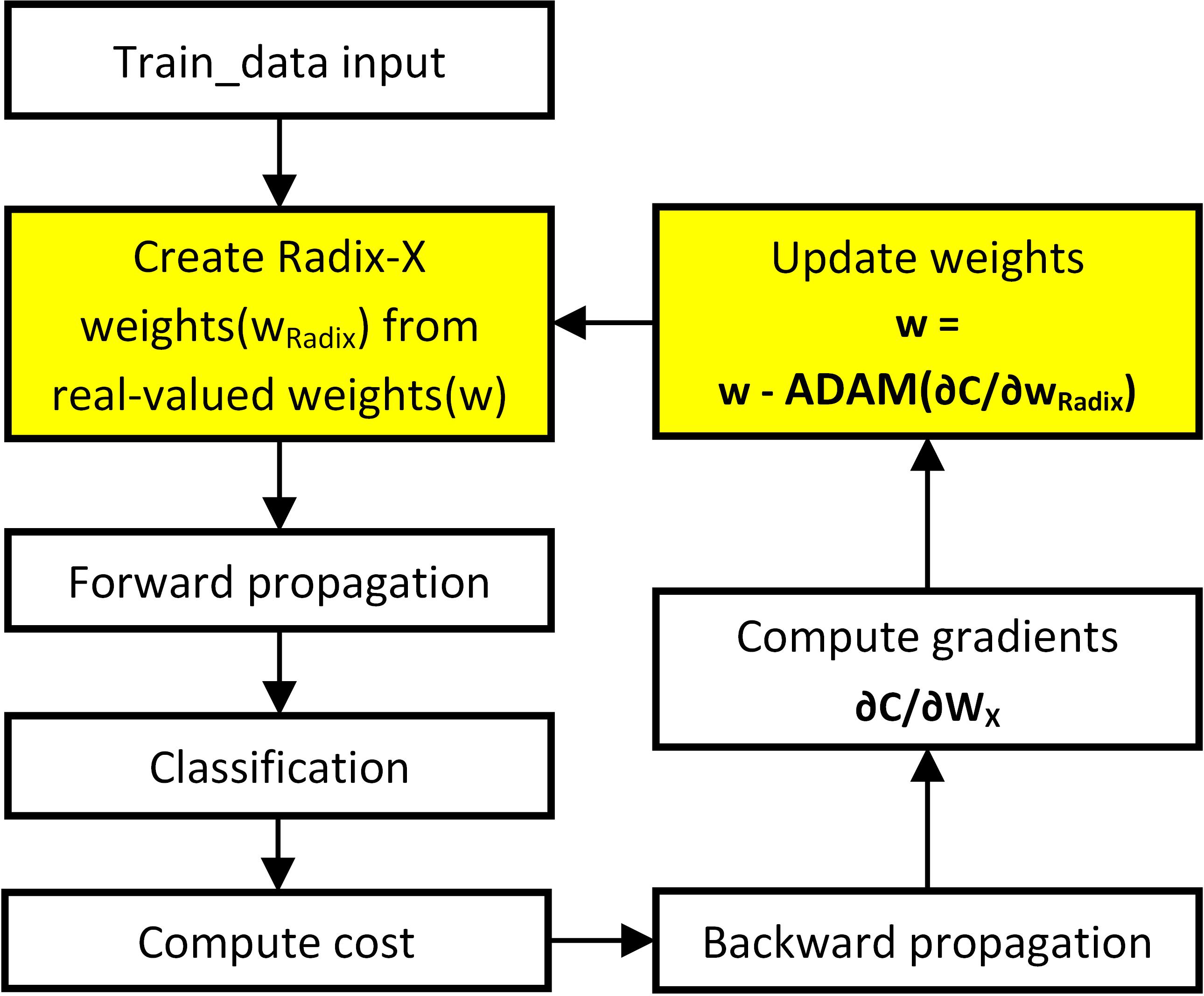}
\caption{The process of training radix-X CNN models. The gradients of the cost function $\delta C / \delta w_{r-X}$ are obtained through forward- and back-propagation using radix-X converted weights $w_{r-X}$. The ADAM optimizer updates real-valued weights $w$ based on those in the previous cycle. This updated $w$ is converted to a radix-X weight $w_{r-X}$ and used as a parameter to decrement the cost function again. The real-valued weight must be saved during training.}
\label{Fig6}
\end{figure}

When training data is passed through the network, the neuron output and weights are converted to $w_{r-5}$ using (\ref{Eq6})-–(\ref{Eq8}). Then, the classification result is obtained through forward propagation. The cost function for the output is obtained, and the slope of the cost function for $w_{r-5}$ is calculated using backward propagation. We compute the real-valued weights using the ADAM optimizer \cite{Kingma2014}, which is used to calculate and store $w_{r-5}$. This feedback process is represented in Fig.~\ref{Fig6}, and while it bears many similarities to conventional backpropagation, we will show how it can be harnessed at the system level using parallel-connected memristive junctions in a crossbar array in the following sections. 

\section{Radix-X CNN Accelerator Circuit}

We designed and fabricated an application specific reconfigurable crossbar array, intended precisely for the implementation of our radix-X CNN Accelerator. Here, we will describe the operating principle of our design, how to achieve multi-bit and negative weights at a single crosspoint, and then detail the nanofabrication techniques used in its development.

\subsection{Multi-bit Weights}
As the resistance precision of the memristor for storing information is limited, and the impact of writing variation increases with the number of resistance states \cite{Ni2016, Stathopoulos2017, Tang2017}, we seek to circumvent this issue by introducing parallel-connected memristors at each crosspoint in the array. Each of these heterogeneous memristors are still only used to store binarized weights, but by forming and severing connections to the memristor electrodes, we introduce additional bits per crosspoint, despite our conservative design approach. 

\begin{figure}
\centering
\includegraphics[width=3.45in]{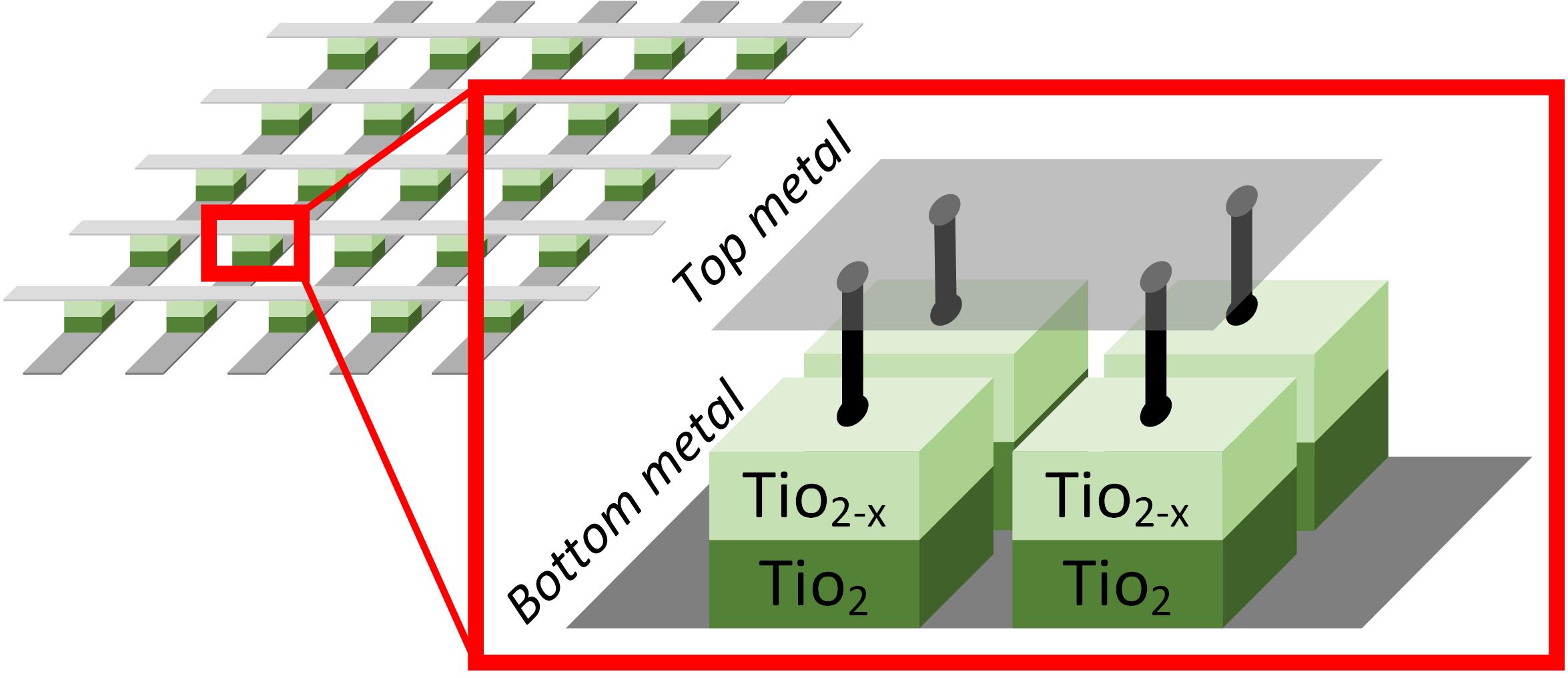}
\caption{Parallel-connected crosspoint for X=5. Each intersection of the crossbar array implements a physically sub-divided memristor into four constituent components. The equivalent circuit is 4-parallel connected memristors that can represent five weight values. The sub-division process is only limited by increased variation in devices of decreasing width.}
\label{Fig7}
\end{figure}

Fig.~\ref{Fig7} depicts an illustration of this concept at a high-level. The crosspoint of a single junction has (X-1) parallel connected memristors. In the diagram shown above, we have chosen X = 5 (i.e., radix-5) which requires four parallel memristors per column-row wire intersection.

\noindent 4 $\times$ 1-bit memristors are placed in a quad-parallel structure per metal crosspoint. Thus, 0-4 memristors are connected to the top metal and pre-programmed to either a HRS or LRS. That is, these memristors are used as read-only memory to ensure high reliability and to avoid write variability.

\begin{figure}
	\centering
\subfloat[]{{\includegraphics[width=1\linewidth]{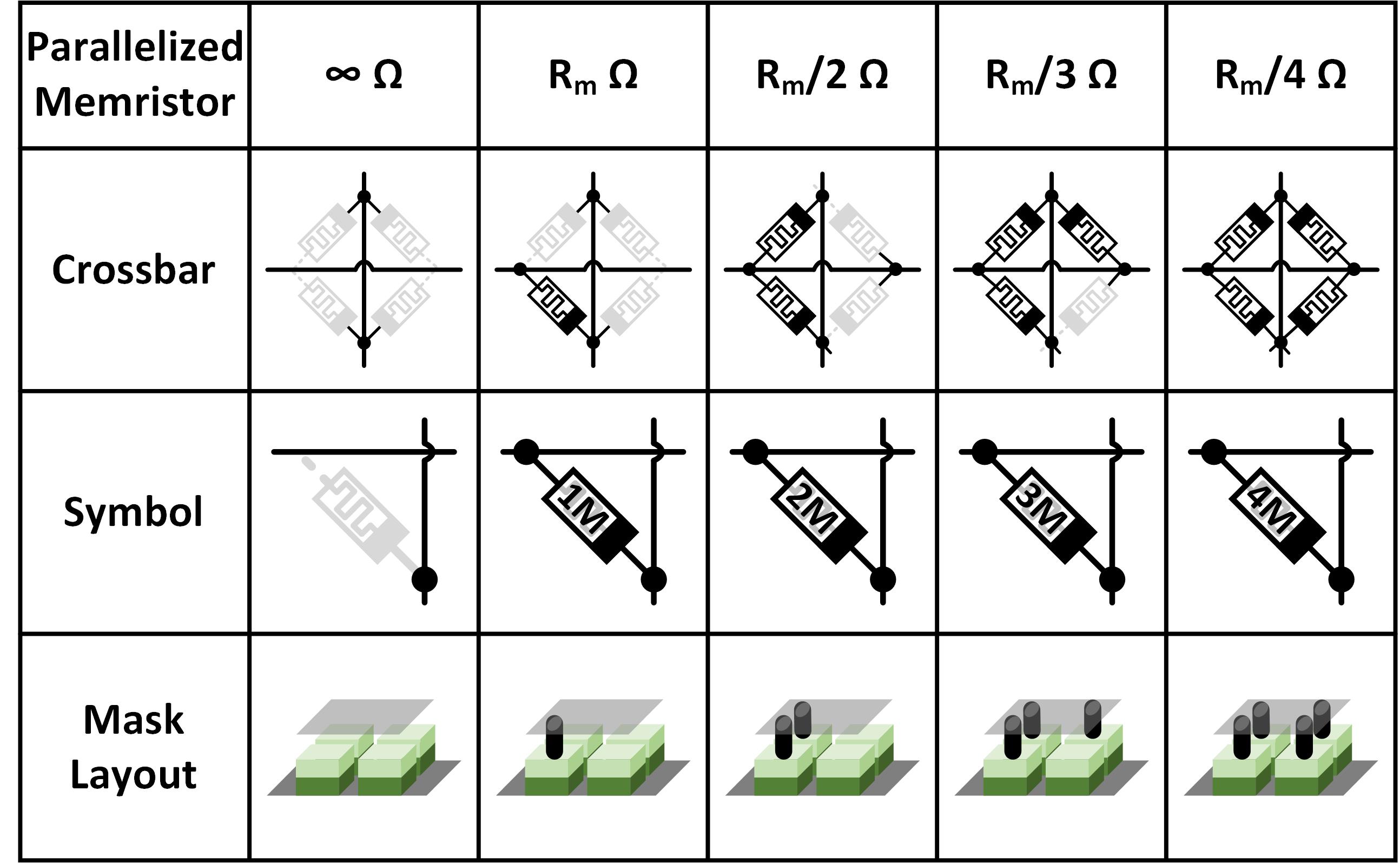}}}
\hfil
\subfloat[]{{\includegraphics[width=0.9\linewidth]{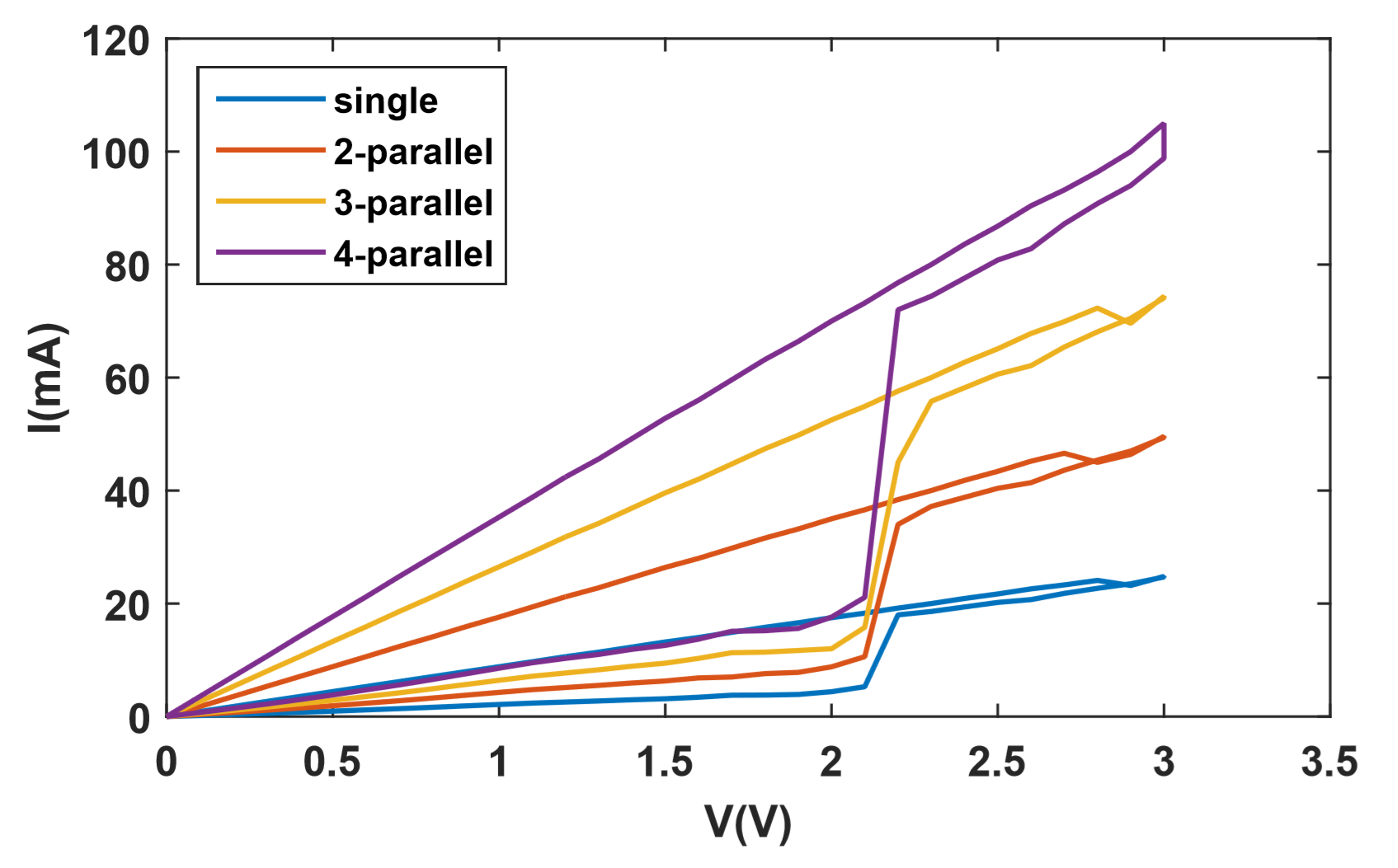}}}
\caption{Parallel-connected memristor crosspoint structure (a) Symbolic resistance, mask layout, number of connections and our shorthand notation to represent each possible type of parallel-connected crosspoint (i.e., 1M to 4M). The mask layout demonstrates that the various parallel connections are hardwired at the time of fabrication, which removes the need for independent access to each memristor within a crosspoint (b) Experimental results obtained by measuring the V-I curves of our fabricated crossbar array from Fig. 12 at frequency = 10Hz.}
\label{Fig8}
\end{figure}

\begin{figure}
	\centering
\subfloat[]{{\includegraphics[width=0.9\linewidth]{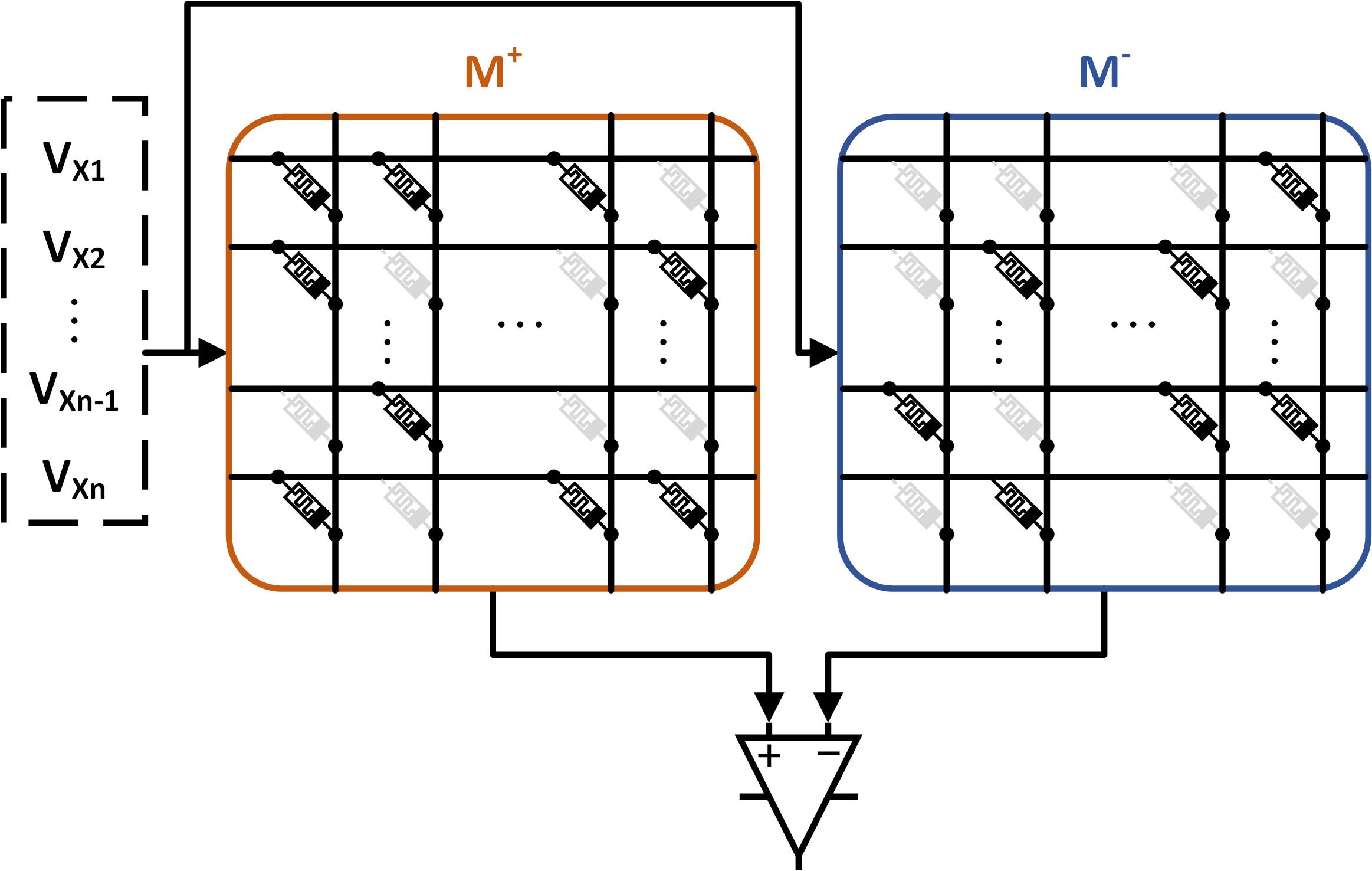}}}
\hfil
\subfloat[]{{\includegraphics[width=0.5\linewidth]{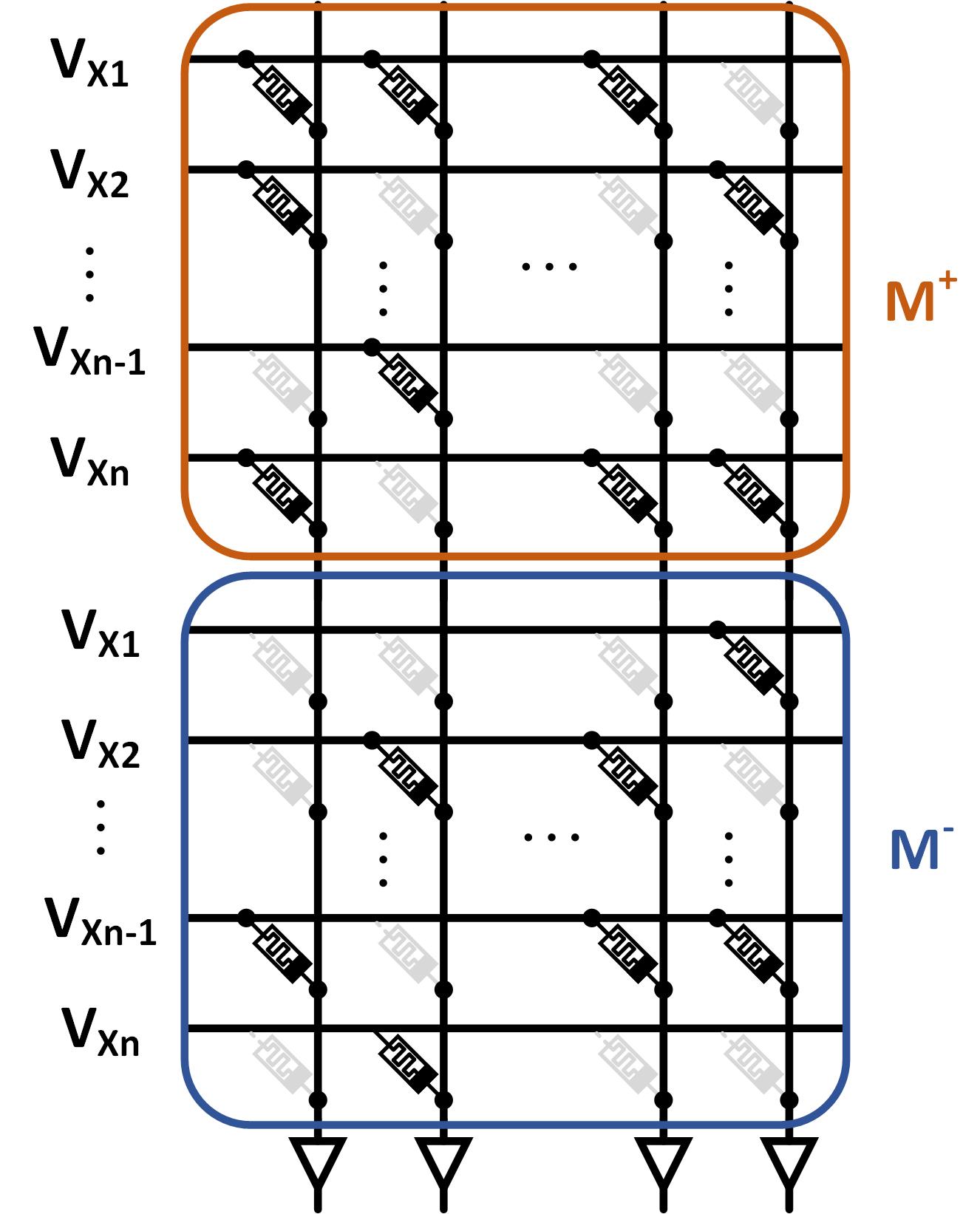}}}
\hfil
\subfloat[]{{\includegraphics[width=0.35\linewidth]{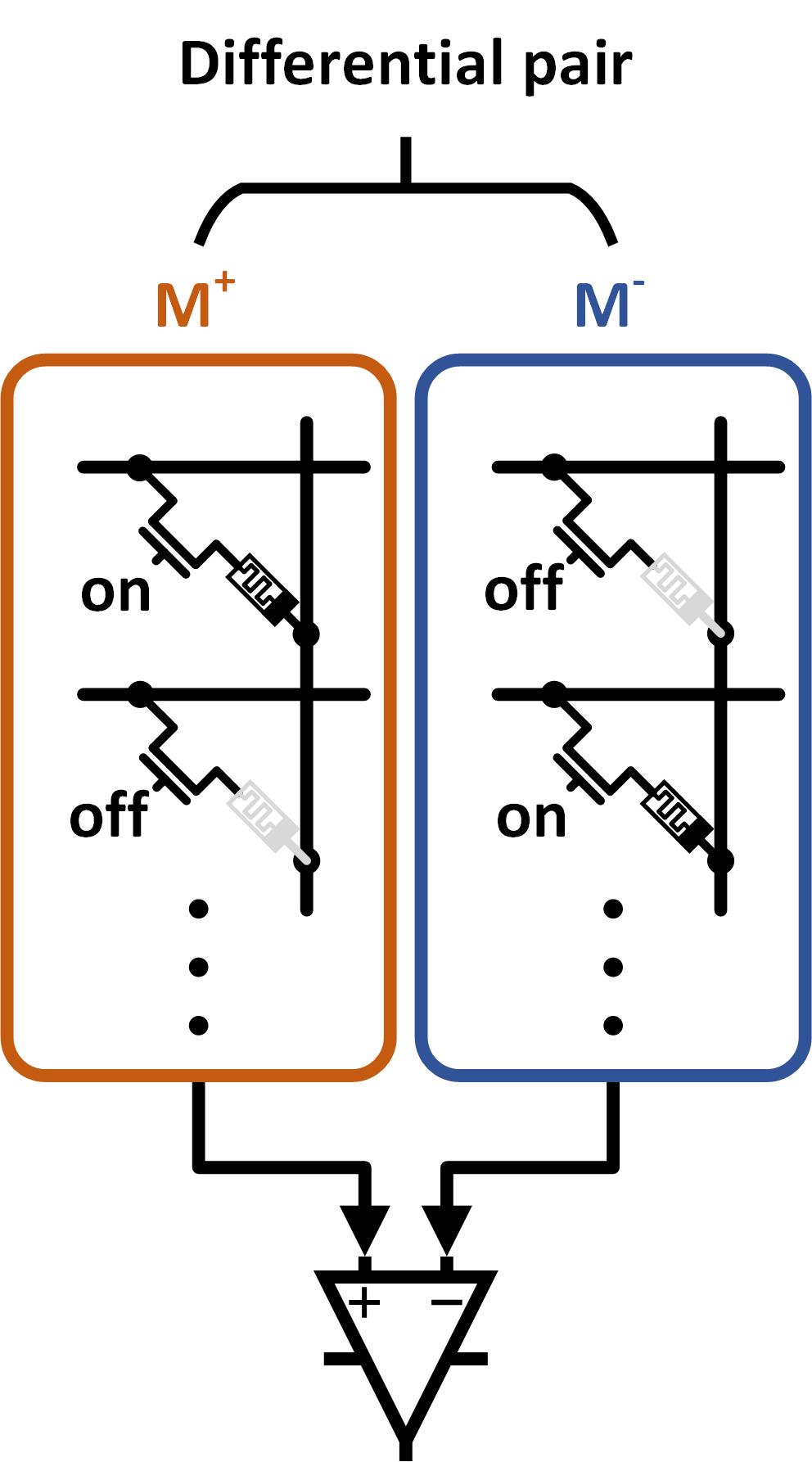}}}
\caption{Negative weights in crossbar arrays (a) Conventional horizontal mapping: array M+ for storing positive weights and M- for storing negative weights. If the value of the weight stored in the memristor is a positive value, the memristor in M+ is activated and in the case of a negative value, the memristor in M- is activated. (b) Vertical mirroring. (c) 1T1M architecture is required to activate or deactivate a memristor.}
\label{Fig9}
\end{figure}

As shown in Fig.~\ref{Fig8}, five resistance values can be obtained depending on the number of activated parallel-connected memristors. The proposed parallel-connected memristor demonstrates how a set of radix-5 CNN weights using 5 discrete resistance states can be implemented.

\subsection{Negative Weights}
Existing studies have implemented the hardware described in Fig.~\ref{Fig9} to represent negative weights, which requires twice the number of columns. In our proposed method, each of the radix-5 weights \{-2, -1, 0, 1, 2\} are mapped to one of five available memristor configurations. This is depicted in Table I. However, in each of the 5 configurations, the equivalent resistance at a crosspoint is still non-negative. We will demonstrate how to remove the need for duplicative columns by mapping negative weights into positive conductances.

\begin{table}[t]
\centering
\caption{Mapping of $w_{r-5}$ to Conductance}
\addtolength{\tabcolsep}{2.5pt}
\begin{tabular}{c c c c c c} 
\hline
\hline \\[-6pt]
\multicolumn{6}{c}{Mapping Values}\\[2pt]
\hline \\[-6pt]
$w_{r-5}$ & -2 & -1 & 0 & 1 & 2\\[5pt]
$G[S]$ & $0/R_m$ &  $1/R_m$ &  $2/R_m$ &  $3/R_m$ &  $4/R_m$\\[1pt]
\hline
\hline \\[-6pt]
\multicolumn{6}{p{\linewidth}}{Mapping five resistances to five conductance to represent all weights of the Radix-5 CNN without duplicative columns. $w_r-5$ = the weights of pre-trained Radix-5 CNN model; G = equivalent conductance of a crosspoint determined by the number of parallel-connected memristors between row and column wires. }
\end{tabular}
\label{Tab1}
\end{table}

First, all radix-X weights $w_{r-X}$ are positively shifted by the magnitude of the minimum weight $w_{r-X,min}$. This translates the minimum weight to 0. Next, each level-shifted weight is divided by the resistance of a single memristor to calculate the equivalent memristance weight. For example, in radix-5, the minimum weight is -2. Where $w_{r-X}$ = -1, a level shift of $|$$-2|$ gives +1, and the equivalent resistance can be found by dividing $R_m$ by this value. Table I shows that only one memristor (1M) should be connected between row and column wires to attain $R_m$. For $w_{R-X}$ = 0, the equivalent resistance will be $R_m/2$. Table I indicates that two memristors are connected in parallel: $1/R_{tot} = (1/R_m + 1/R_m) \implies R_{tot} = R_m/2$. The equivalent conductance is given by:

\begin{equation} \label{Eq9}
G = \frac{{w_{r-X}}+|w_{r-X,min}|}{R_m}.
\end{equation}

\noindent Substituting (\ref{Eq9}) into (\ref{Eq2}) gives the following equation for the column current for $n$ rows in radix-5:

\begin{equation} \label{Eq10}
i_{tot} = \sum_{i=0}^{n} \Big(\frac{w_{r-5,i}+2}{R_m}*V_i \Big).
\end{equation}

\begin{figure}
	\centering
\subfloat[]{{\includegraphics[width=0.45\linewidth]{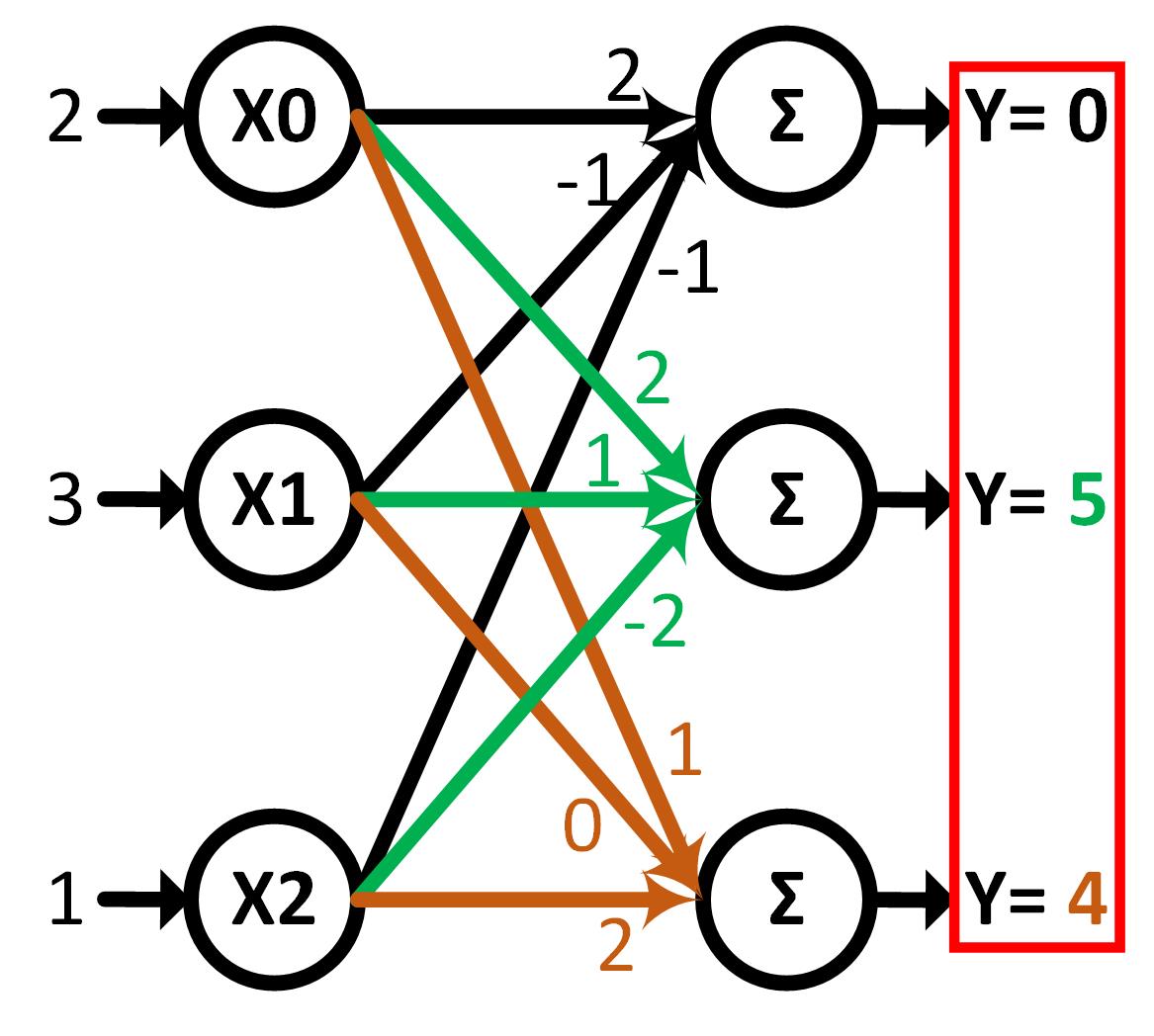}}}
\hfil
\subfloat[]{{\includegraphics[width=0.45\linewidth]{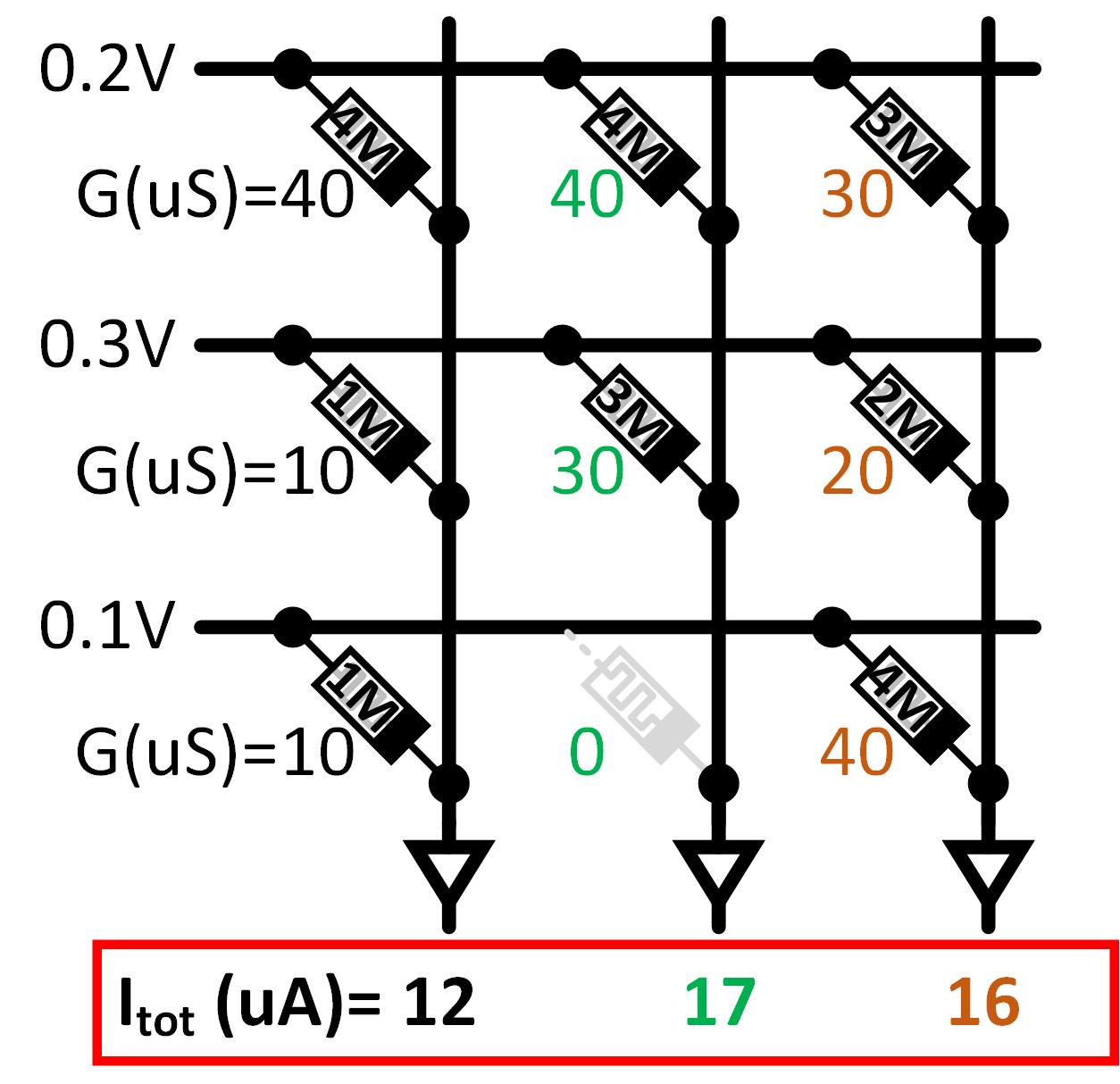}}}
\hfil
\subfloat[]{{\includegraphics[width=0.45\linewidth]{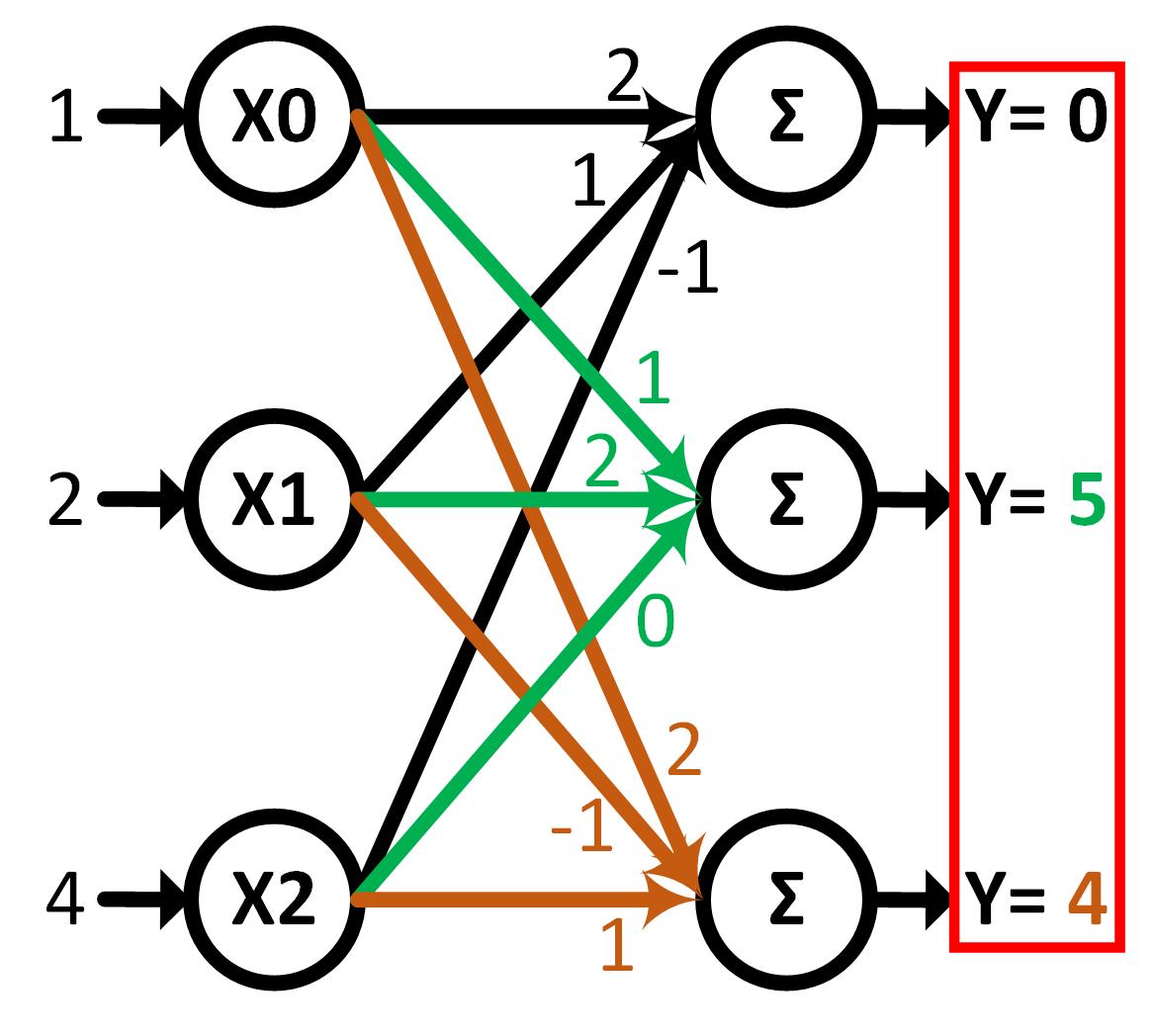}}}
\hfil
\subfloat[]{{\includegraphics[width=0.45\linewidth]{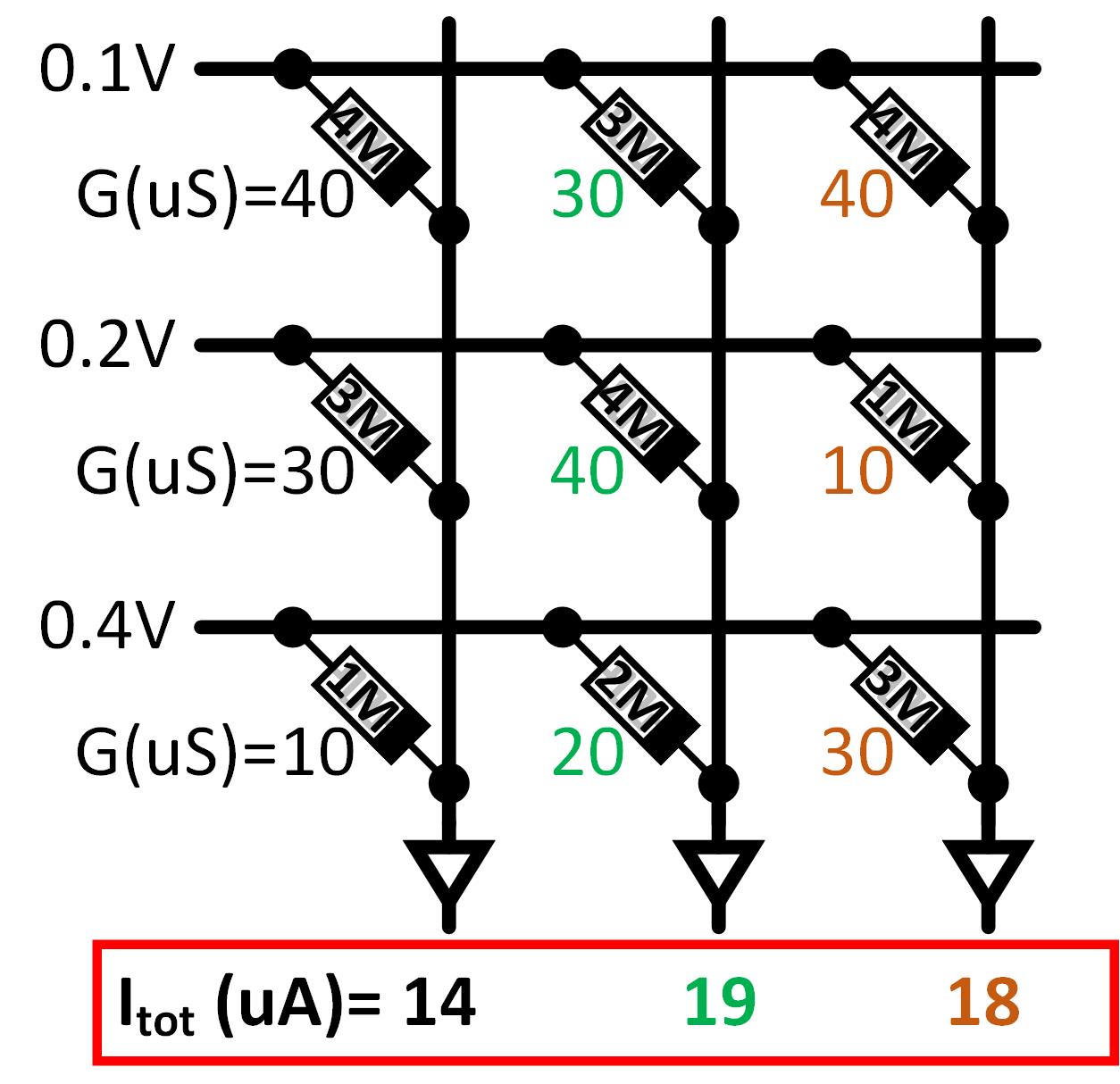}}}
\caption{A simple example showing the issues that occur when linearly mapping negative weights to positive conductances in an ANN without a level-shift: $R_m~=~100~\text{k}\Omega$, $V_i = x/10$ V. All values are color co-ordinated.}
\label{Fig10}
\end{figure}

\noindent However, this is an insufficient representation of output current. To see why, consider Figs.~\ref{Fig10}(a) and (c) which are radix-5 ANNs consisting of 3 unique inputs, and Figs.~\ref{Fig10}(b) and (d) which are the crossbar array equivalents using our parallel-connected structure. The equivalent conductances are derived from Table I and (\ref{Eq9}), where $R_m = 100~\text{k}\Omega$. The current through the first column in Fig.~\ref{Fig10}(b) is calculated using (\ref{Eq10}): 

\begin{equation} \label{Eq11}
i_{tot} = \frac{2+2}{100k}*0.2 + \frac{-1 + 2}{100k}*0.3 + \frac{-1 + 2}{100k} * 0.1 = 12 \mu A,
\end{equation}

\noindent and for the first column of Fig.~\ref{Fig10}(d):

\begin{equation} \label{Eq12}
i_{tot} = \frac{2+2}{100k}*0.1 + \frac{1 + 2}{100k}*0.2 + \frac{-1 + 2}{100k} * 0.4 = 14 \mu A.
\end{equation}

\noindent Although the outputs of the two ANNs in Figs.~\ref{Fig10}(a) and (c) are identical, the read-out currents from Figs.~\ref{Fig10}(b) and (d) are different. This is a result of element-wise level shifts of weights causing subsequent mismatch. 

To counter the $w_{r-X,min}$ level-shift, we must design an adaptive reference line to be subtracted from the signal columns. To do this, we note that the minimum column current $12\mu A$ in Fig.~\ref{Fig10}(b) corresponds to the ANN output of Y = 0. If we subtract $12\mu A$ from each current in the set \{$12\mu A$, $17\mu A$, $16\mu A$\}, the resulting set of column currents becomes \{$0\mu A$, $5\mu A$, $4\mu A$\}; there is now a 1:1 correspondence to the ANN output. For Fig.~\ref{Fig10}(c), subtracting the minimum current $14\mu A$ from \{$14\mu A$, $19\mu A$, $18\mu A$\} attains \{$0\mu A$, $5\mu A$, $4\mu A$\}. The current sets now match the ANN outputs. In both cases, the solution is to subtract the current corresponding to the ANN output of `0' from all column signals.

In a radix-5 crossbar array, we create our own zero-weight reference column by having two memristors in parallel at each row (2M in Table I). This corresponds to a radix-5 weight of 0 for an entire column.\footnote{This is generalizable beyond radix-5 to radix-X, where the zero-weight conductance can be calculated by substituting $w_{r-X}=0$ into (\ref{Eq9}). It is this ability to generalize that enables our algorithm to have an adaptive precision: radix-5 is simply a test case for demonstration.}  The output current of the reference line can be calculated by substituting $w_{r-5} = 0$ into (\ref{Eq10}):

\begin{equation} \label{Eq13}
i_{ref} = \sum_{i=0}^{n} \Big(\frac{2V_i}{R_m}\Big).
\end{equation}

\noindent This is generalized to any radix-X numeral system, by substituting $w_{r-X} = 0$ into (\ref{Eq9}), and the result into (\ref{Eq2}):

\begin{equation} \label{Eq14}
i_{ref} = \sum_{i=0}^{n} \Big(\frac{|w_{r-X,min}|}{R_m}*V_i\Big).
\end{equation}

\noindent The reference current is dependent on the input voltages, and therefore cannot be implemented using a constant current. This was demonstrated by example in Fig.~\ref{Fig10}. The reference current $i_{ref}$ is converted into a voltage using an op-amp, and subtracted from all signal voltages with an array of differential amplifiers.

\begin{figure}
\centering
\includegraphics[width=3in]{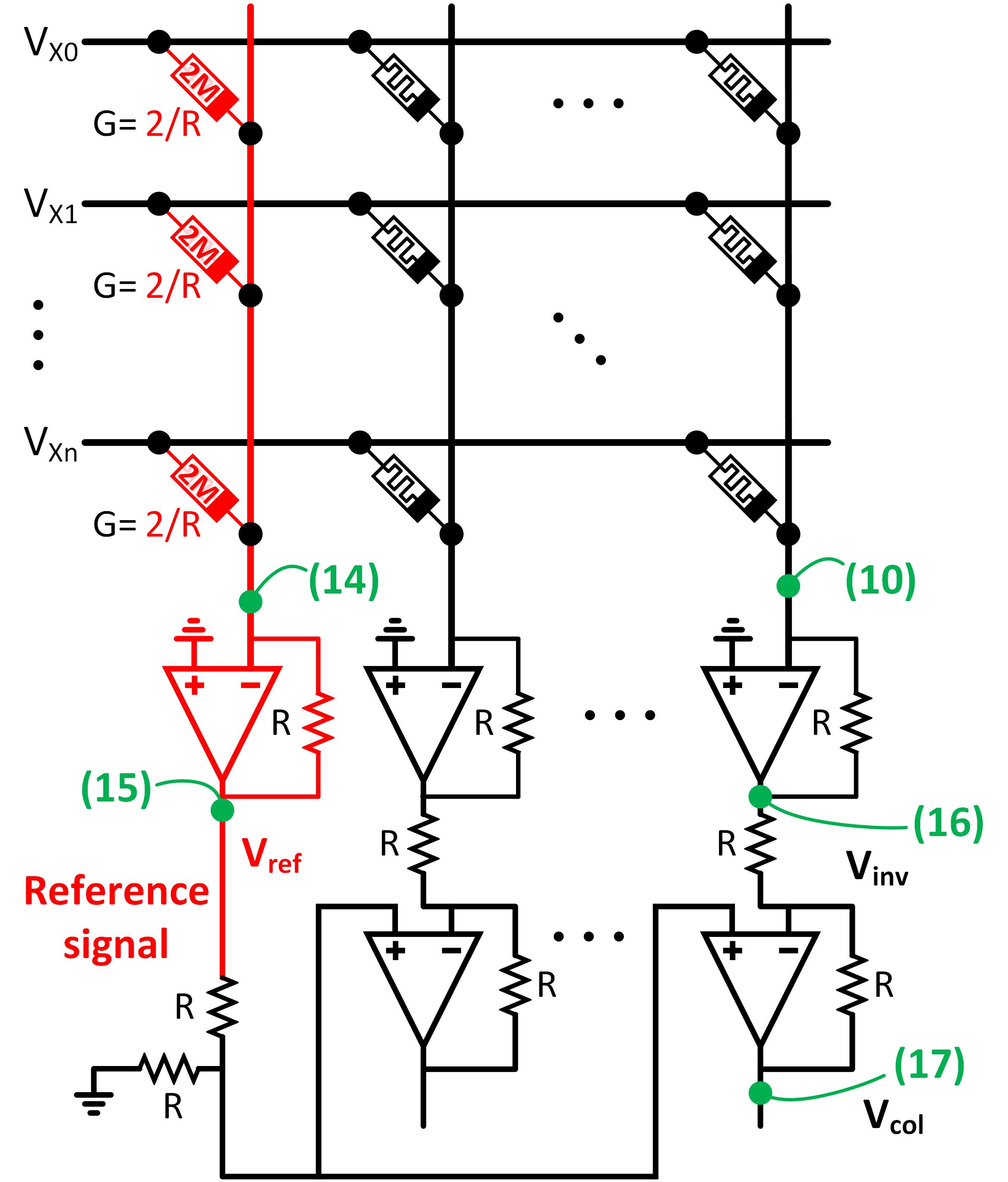}
\caption{The structure of the proposed circuit is able to represent negative weights with the addition of a single column. The linear shift of the reference signal is applied by subtracting $V_{ref}$ from the output of each column. $V_{inv}$ is the output of the inverting amplifier and $V_{col}$ is the output voltage of a single column, and corresponds to an artificial neuron. The relevant equation number for each signal is shown in the figure.}
\label{Fig11}
\end{figure}

The hardware level implementation of the level-shift is shown in Fig.~\ref{Fig11}, with the reference line highlighted in red. The inverting amplifiers are used to fix all columns at virtual ground. To find the potential at the output of the inverting amplifier on the reference line, note that $i_{ref}$ from (13) is passing through the negative feedback resistor $R$:

\begin{equation}  \label{Eq15}
\begin{split}
V_{ref} & = -R * i_{ref} \\
&=-R*\sum_{i=0}^{n} \Big(\frac{2V_i}{R_m}\Big) \because \text{(13)}.
\end{split}
\end{equation}

\noindent Similarly, for the inverting amplifier output of the signal columns: 

\begin{equation}  \label{Eq16}
\begin{split}
V_{inv} & = -R * i_{tot} \\
&=-R\bigg[\sum_{i=0}^{n} \Big(\frac{w_{r-5,i}+2}{R_m}*V_i\Big)\bigg] \because \text{(10)}.
\end{split}
\end{equation}

\noindent Given all resistors of the differential amplifier are equivalent, the output stage of the crossbar array is a subtractor with $V_{ref}$ from (\ref{Eq15}) being passed into the positive terminal, and $V_{inv}$ from (\ref{Eq16}) into the negative terminal:

\begin{equation}  \label{Eq17}
\begin{split}
V_{col} & = V_{ref} - V_{inv} \\
&=R\sum_{i=0}^{n} \Big(\frac{w_{r-5,i}+2}{R_m}*V_i\Big)-R\sum_{i=0}^{n} \Big(\frac{2V_i}{R_m}\Big)\\
&=R*\sum_{i=0}^n\Big(\frac{w_{r-5,i}*V_i}{R_m}\Big).
\end{split}
\end{equation}

\noindent The final result of (\ref{Eq17}) shows how the `+2' linear shift is removed by $V_{ref}$, thus ensuring a correct representation of negatively weighted MVMs following the demonstration in Fig.~\ref{Fig10}.

The relationship between a neural network input $X_n$ and the input voltage $V_n$ in the circuit is given as,

\begin{equation}\label{Eq18}
V_n = \frac{X_n}{S},
\end{equation}

\noindent where $S$ is the scaling factor of $V_{n}$. Substituting (\ref{Eq18}) and (\ref{Eq1})into (\ref{Eq17}) obtains:

\begin{equation}  \label{Eq19}
\begin{split}
V_{col} & = R * \sum_{i=0}^n\Big(\frac{w_{r-5,i}*X_i}{R_m*S}\Big) \\
& = \Big(\frac{R}{R_m*S}\Big)*Y \because \text{(1)}, \text{(17)}.
\end{split}
\end{equation}

\noindent This verifies that the output voltage of our radix-X CNN accelerator is simply scaled by $(\frac{R_m*S}{R})$, and concludes that we are able to represent multi-bit negative weights with a parallel-connected memristor without duplicative columns.

\section{Simulation Results}

\begin{figure}
\centering
\includegraphics[width=3in]{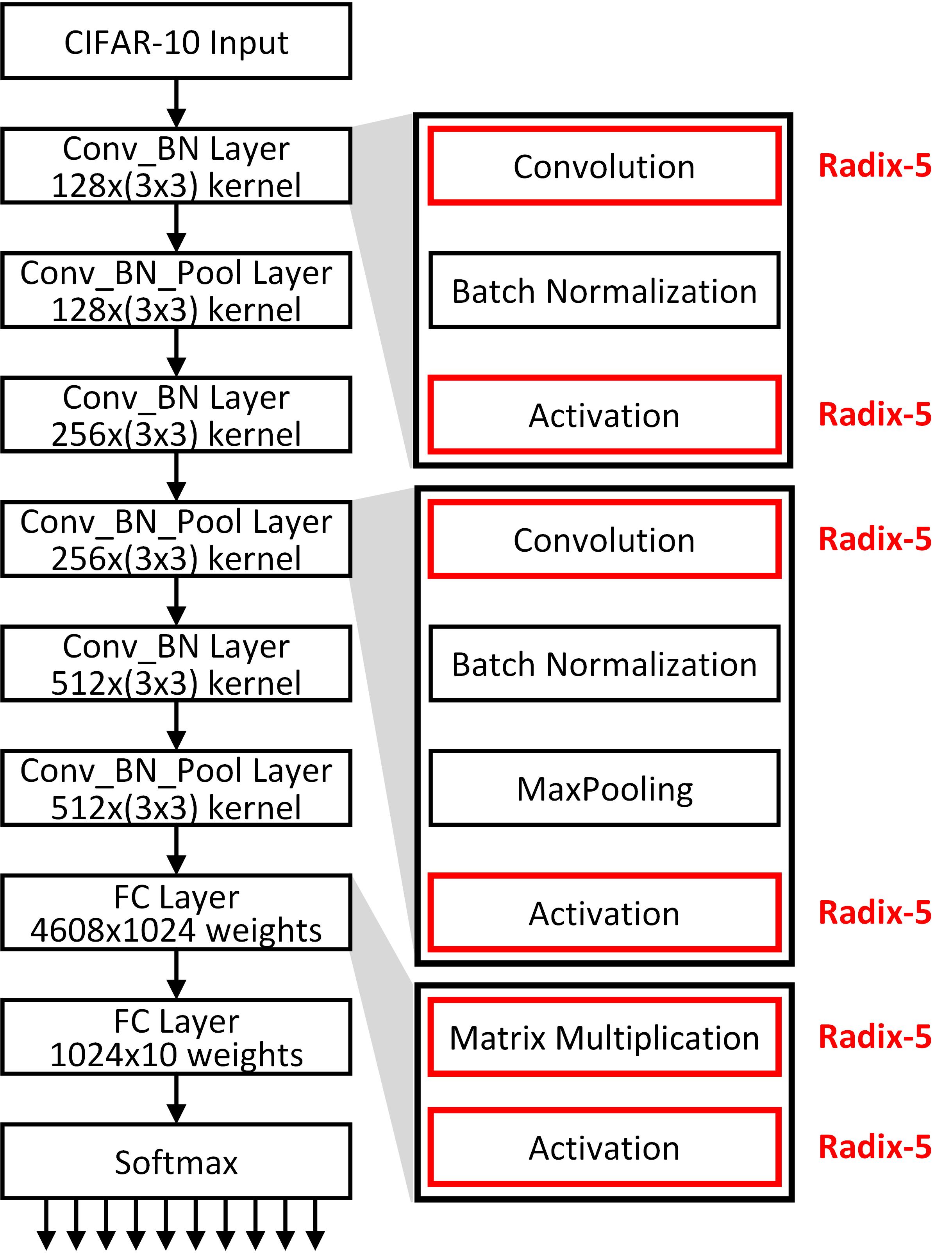}
\caption{The architecture of the CNN used. Radix-5 weight and activation blocks are marked in red.}
\label{Fig12}
\end{figure}

\begin{figure}
\centering
\includegraphics[width=3in]{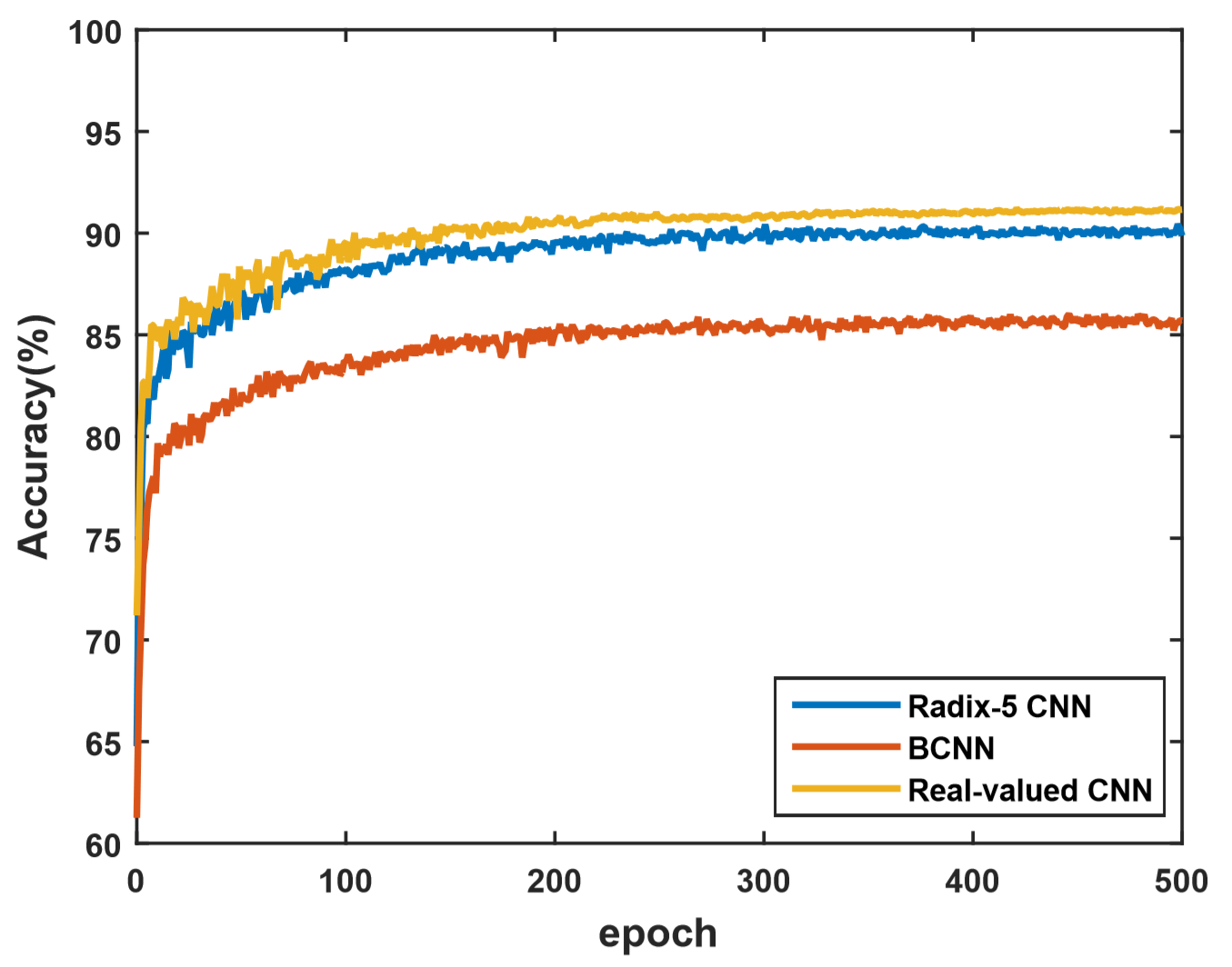}
\caption{Validation accuracy of a real-valued CNN, BNN and the proposed radix-5 CNN. While the training performance of the CNN using full precision weight and radix-5 CNN shows a slight difference, the use of a BNN results in a noticeable drop in performance.}
\label{Fig13}
\end{figure}

We conducted a simulation of the radix-X CNN accelerator described above with all memristors being used as read-only memory, and peripheral circuitry in the SK Hynix 180nm CMOS Process. The characteristics of the simulated memristor are based on our own \ce{Al/TiO2}/\ce{TiO_x/Al} crossbar array which we will provide details of in the next section. The relevant features for our feed-forward simulation of a pre-trained network are $R_m = 100 \text{k}\Omega$ and $V_{Th} = 0.5 V$. As all parallel configurations are  fixed on our crossbar, there was no need to consider switching time characteristics and programming variations. The relatively large width of our metal lines ($20 \mu m$) meant low line resistance and so line losses were negligible. When scaling the metal lines down and the number of rows and columns up, this assumption will need to be adapted accordingly. The final idealization made was assuming negligible device-to-device variation which was accounted for in experimentation. The peripheral resistances were chosen to be $R = 10 \Omega$ and the scaling factor $S = 10$ to ensure read voltages did not exceed the switching threshold.

\begin{table}[t]
\centering
\caption{Comparison of Memristor Based CNNs for CIFAR-10}
\begin{tabular}{c c c c c} 
\hline
\hline \\[-6pt]
 & \multicolumn{2}{c}{\underline{Training Accuracy}} & Validation & Area*\\
 & 10 epochs & 500 epochs & Accuracy & ($\mu m^2$)\\
\hline \\[-6pt]
\textit{Real-valued CNN} & 92\% & 99\% & 91.5\% & 8400\\
\textit{BNN} & 88\% & 99\% & 86.0\% & 8400\\
\textit{Radix-5 CNN} & 92\% & 99\% & 90.5\% & 4600\\
\hline
\hline \\[-6pt]
\multicolumn{5}{p{\linewidth}}{*SK Hynix 180nm CMOS process. Area is based on the layout in the BEOL before the pad level, where CNN and BNN implementations require differential pairs for signed weight representation from Fig.~\ref{Fig9}.}
\end{tabular}
\label{Tab2}
\end{table}

The architecture of our radix-5 CNN is shown in Fig.~\ref{Fig12}. We evaluated the validation accuracy for three implementations of a high precision 16-bit CNN, BNN and the proposed radix-5 CNN. Fig.~\ref{Fig13} shows the classification accuracy during training on the CIFAR-10 dataset, where a high precision CNN and radix-5 CNN showed a difference in accuracy of approximately 0.8\%. This is a 5.3\% improvement over BNNs, which this is to be expected given the higher base value used, but for a substantial decrease in chip area. A more detailed comparison is summarized in Table II.

\begin{figure}
	\centering
\subfloat[]{{\includegraphics[width=0.5\linewidth]{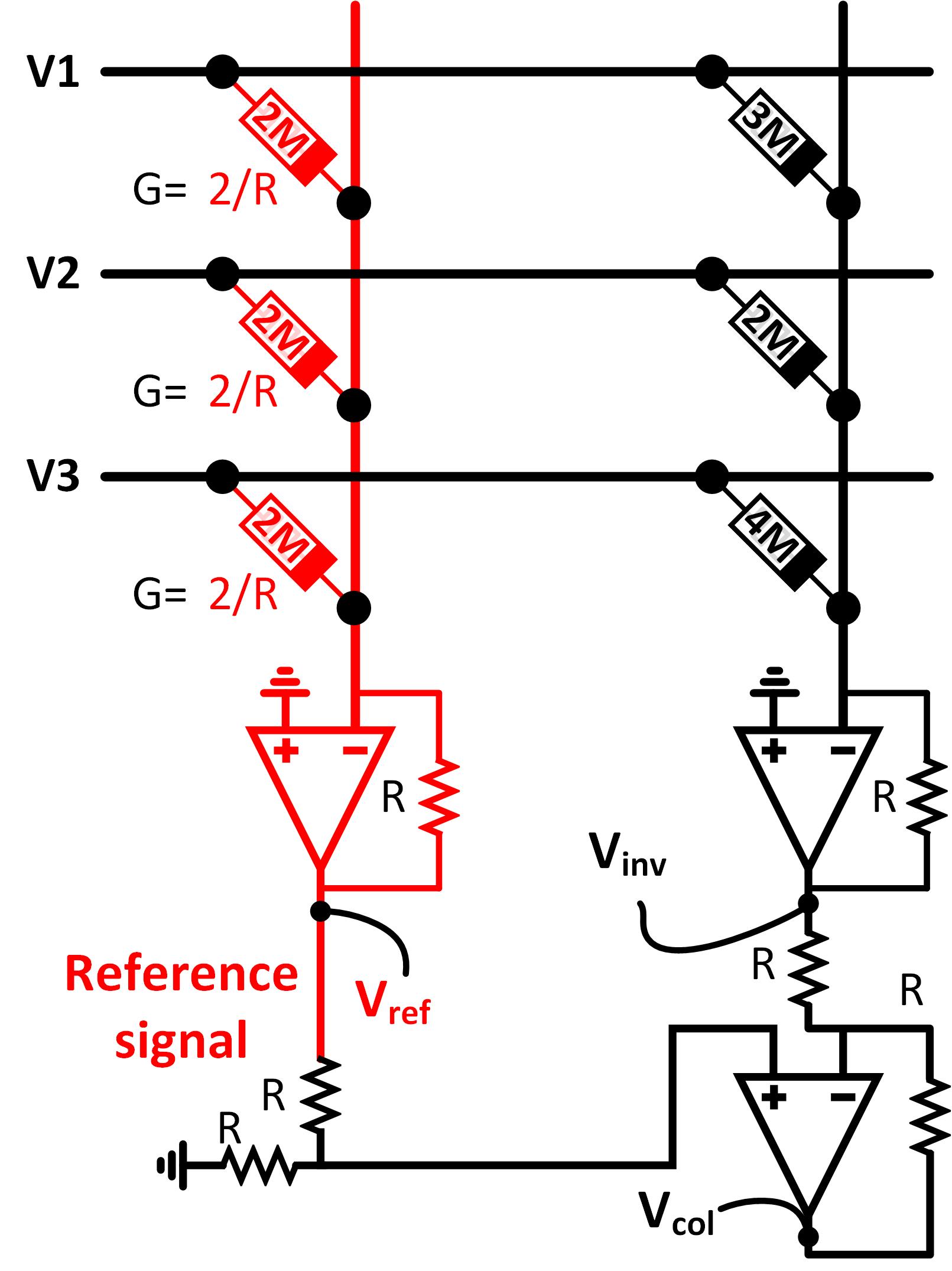}}}
\hfil
\subfloat[]{{\includegraphics[width=0.95\linewidth]{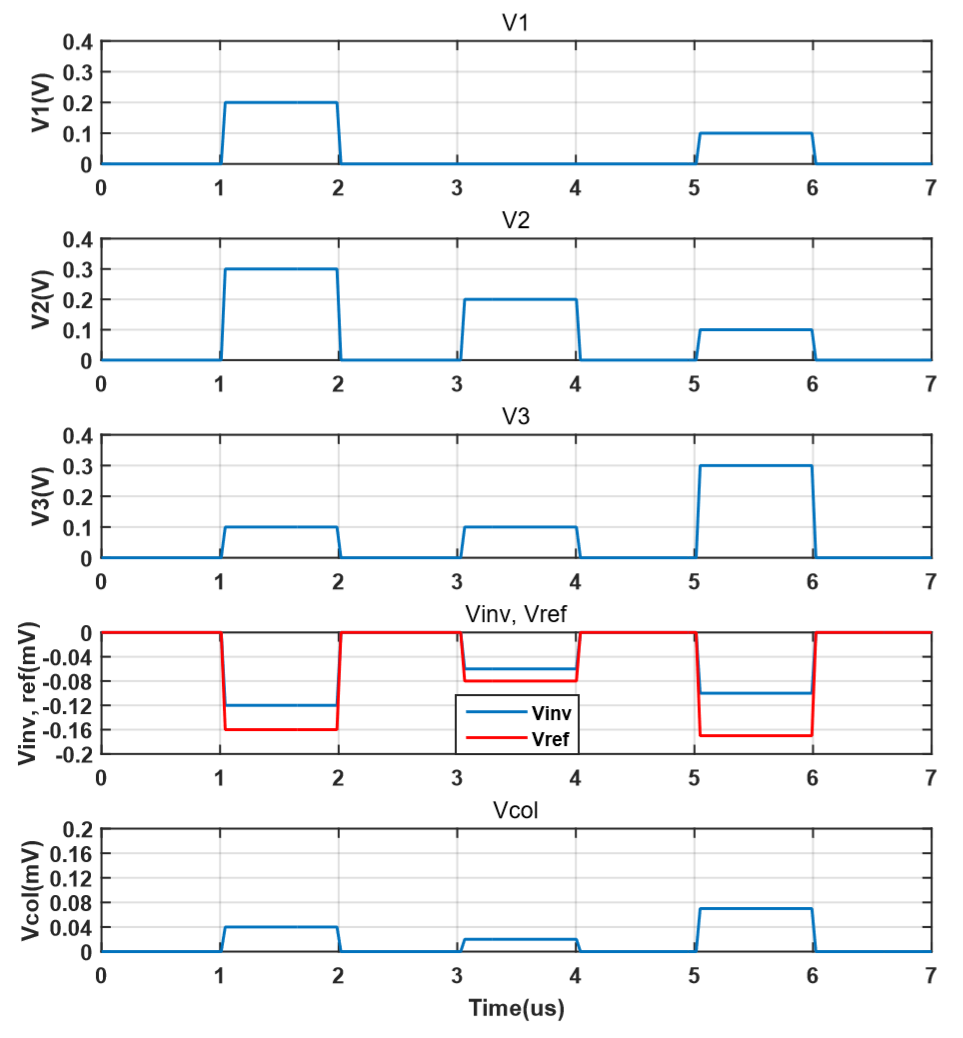}}}
\caption{(a) Simulation of the proposed architecture where $R_m = 100 \text{k}\Omega$, $R = 10 \Omega$, and $S = 10$. (b) Output of each node for various inputs.}
\label{Fig14}
\end{figure}

As shown in Fig.~\ref{Fig14}, the behavior of a simple neural network for the proposed radix-5 CNN is  fully implemented and simulated. Analyzing the simulation results in Fig.~\ref{Fig14}(b) shows that the output of the neuron for the first pulse during time $t = 1 \mu s$ to $2 \mu s$ is verified with (\ref{Eq2}):

\begin{equation} \label{Eq20}
Y = 2*1+3*0+1*2 = 4 \because (1).
\end{equation} 

\noindent Therefore,

\begin{equation}\label{Eq21}
V_{col} = \Big(\frac{10}{100k*10}\Big)*4 = 40 \mu V \because (19).
\end{equation}

\noindent In the same manner as (\ref{Eq20}) and (\ref{Eq21}), the $V_{col}$ outputs from the second and third input pulses are $20 \mu V$ and $70 \mu V$, respectively. The results of our simulation agree with our mathematical derivations in section IV.

\section{Experimental Results}
\subsection{Nanofabrication}

\begin{figure}
\centering
\includegraphics[width=3.45in]{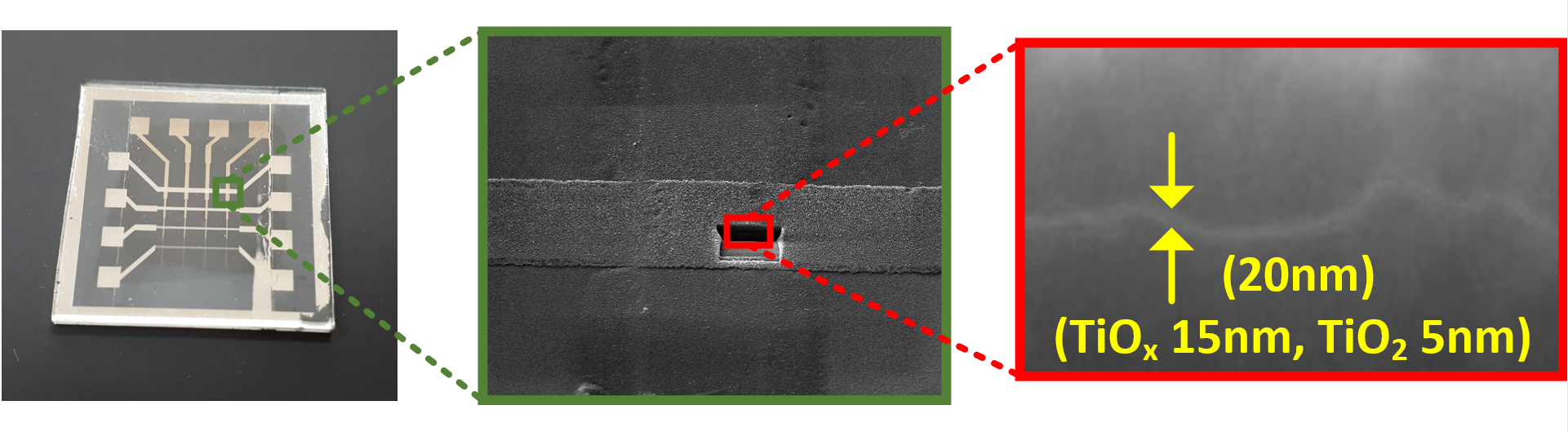}
\caption{\ce{Al/TiO2}/\ce{TiO_x/Al} memristor device imaged using a focus ion beam (FIB) analyzer.}
\label{Fig15}
\end{figure}

We fabricated a proof-of-concept $4 \times 4$ parallel-connected crossbar array in-house to demonstrate the feasibility of the proposed memristor-based radix-5 CNN method. This  was achieved with a sandwich structure composed of \ce{Al/TiO2}/\ce{TiO_x/Al} layers. A 200-nm-thick Al layer was deposited as the bottom electrode on a glass wafer. Standard photolithography was conducted to produce 20-$\mu$m-wide Al lines. During the microfabrication process, the wafers were irradiated by using a mask alignment system for 100~s and then developed at 296K for 120~s. The Al channel was then defined by wet etching (\ce{H3PO4}:\ce{HNO3}:\ce{CH3COOH}:\ce{H2O} = 80 ml~:~5 ml~:~5 ml~:~10 ml), removing any Al outside of the channel regions at an etching rate of $\Delta d/t$ = 300 nm/min. 5-nm-thick \ce{TiO2} thin film and a 15-nm-thick \ce{TiO_x} thin film layers were formed by atomic layer deposition (ALD) and magnetron sputtering. Subsequently, another 200-nm-thick Al layer was sputtered as the top electrode, followed by standard photolithography to create $20 \mu m \times 20 \mu m$ windows. Fig.~\ref{Fig15} shows a cross-sectional image of a single memristor taken with a focus ion beam (FIB) analyzer.

\subsection{Image Processing}

\begin{figure}
	\centering
\subfloat[]{{\includegraphics[width=0.47\linewidth]{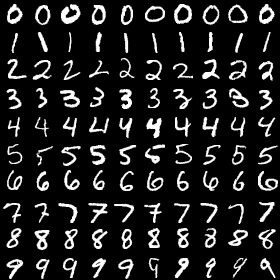}}}
\hfil
\subfloat[]{{\includegraphics[width=0.47\linewidth]{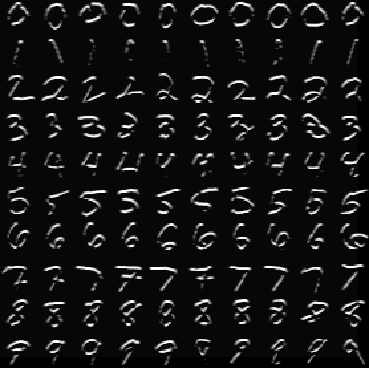}}}
\caption{2D convolution performed on 100 $28 \times 28$ images from the MNIST dataset with a Sobel filter in radix-5 on the crossbar array in Fig.~\ref{Fig15}. (a) Before processing. (b) After processing.}
\label{Fig16}
\end{figure}

\begin{figure}
	\centering
\subfloat[]{{\includegraphics[width=0.3\linewidth]{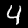}}}
\hfil
\subfloat[]{{\includegraphics[width=0.3\linewidth]{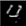}}}\\
\subfloat[]{{\includegraphics[width=0.3\linewidth]{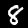}}}
\hfil
\subfloat[]{{\includegraphics[width=0.3\linewidth]{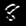}}}
\caption{A scaled-up inspection of hardware-based 2D convolution. (a) Before processing digit `4'. (b) After processing digit `4'. (c) Before processing digit `8'. (d) After processing digit `8'.}
\label{Fig17}
\end{figure}

We performed image convolution on 100 images of handwritten digits from the MNIST dataset, of $28 \times 28$ in dimension \cite{LeCun1998} and passed them through a Sobel filter, which is typically used in edge detection algorithms. The Sobel operator takes the form of a $3 \times 3$ matrix in radix-5 form: 

\begin{equation}\label{Eq22}
\textbf{G}=\begin{bmatrix} 
1 & 2 & 1\\
0 & 0 & 0\\
-1 & -2 & -1
\end{bmatrix}
\end{equation}

\noindent The rationale being that, if the crossbar is capable of performing MVMs then by extension, classification tasks using a CNN will also be possible on larger arrays. The image is processed using similar parameters to those in the simulations, where input pixels are linearly mapped from a null input for a black pixel and $0.4~V$ for a white pixel. As per Table I, a kernel element of `-2' is implemented as an open junction at a crosspoint, and an element of `2' mapped to four parallel connected memristors. The maximum current drawn from a memristor was measured to be approximately $1.6\mu A$, and the critical value for $i_{tot}$ from a full column under the test case of MNIST images passed through an edge detection filter was $4.0 \mu A$. This column current is relatively small when compared to similar  arrays based on conduction via oxygen vacancies, but this is a result of having a small-scale array rather than low read voltages. The output voltages at $v_{col}$ were then linearly mapped back into output images. Qualitatively, we successfully generated a near perfect 2D convolution with a stride of 1 and no zero-padding, as can be seen in Fig.~\ref{Fig16}, and a scaled up sample in Fig.~\ref{Fig17}. The small scale prototyped nature of our array meant that for a $3 \times 3$ kernel, each pixel required 3 read cycles where 4 output pixels could be pipelined across columns, and convolving a $28 \times 28$ image required a total of 21 read cycles.

\section{Discussion}
Implementing BNNs on memristor crossbars is a common technique used to enhance robustness of crossbar arrays in light of analog write variability. Our proposed technique follows this conservative design methodology where the radix-X CNN accelerator uses single-bit memristors. Rather than using binarized encoding across multiple columns, we instead modulate the number of memristors at crosspoints between row and column lines (i.e., a 1T\textit{X}M cell), and have thus proposed a new crossbar architecture and co-developed an algorithm specifically suited to adapt to the number of memristors per cell. 
The first trade-off to consider is the number of additional memristors per cell, as against additional columns to improve precision and implementation of negative weights. This analysis is process dependent, and in our array where the metal lines occupy a width of 20 microns, the minimum width of a single memristor is of sub-micron pitch (and of a few nanometers in more advanced processes \cite{Pi2019, Zhu2019}). For single-bit memristors in conventional binarized crossbars, the closest equivalent comparison to radix-5 is by using 2-bit weights, which will require a total of 4 columns (2 for positive weights, and 2 for the differential pair). We are able to implement the above scheme in 2 columns, with a 20\% improvement in precision using radix-5 over 2-bit representations. The alternative option for column reduction is to use analog weights, which remains a developing but promising field of research. The limiting factor is where the radix of the numeral system becomes larger, resulting in an increasing number of parallel-connected memristors per cell, and an associated reduction in equivalent resistance. Larger metal lines and more vias are needed to cope with the increasing current capacity. While our array had no issues with a critical current of $4.0 \mu A$ (due to the wide metal lines used in our process, and had current capacity of over $100mA$ -- see Fig.~\ref{Fig8}(b)), this will become an increasingly important trade-off when optimizing for higher values of X in radix-X. The effect of decreasing equivalent resistance can be partially mitigated by reducing the read voltage, where state-of-the-art crossbar arrays have demonstrated read currents of under $10nA$ \cite{Fuller2019}.

The second trade-off is with respect to pipelining. Given that parallel-connections are fixed at the time of fabrication, the radix-X crossbar will typically be optimized for specific conductance matrices. In general, this will be advantageous only for kernels containing a particular set of elements. The benefit to reduced reconfigurability is that write-variability is no longer an issue, and endurance is also prolonged due to the application of only read pulses. 

\section{Conclusion}
We have proposed a crossbar array with multiple metal-oxide thin film switches at each crosspoint, and a co-designed algorithm tailored for this inference accelerator to convert a set of pre-trained weights into values based on user-selected precision. We conducted CNN classification on the CIFAR-10 dataset using a large-scale simulation, and performed experimental validation of convolution image processing on a subset of the MNIST dataset using a small-scale crossbar array. We demonstrated that we could achieve multi-bit and negative weights using 46\% of the area of conventional differential pairs of columns, all whilst including an adaptive precision mechanism within our array. What has been proposed is not an exhaustive use of this array. For example, future work includes the use of transistor switches to reconfigure the number of memristors at each crosspoint to enable a higher degree of reconfigurability. Alternatively, as research on multi-bit memristors matures and values of memristance increases, these will be the proponents to achieving higher precision by extending the range of possible base values usable for a given crossbar dimension in radix-X.

\begin{IEEEbiography}[{\includegraphics[width=1in,height=1.25in,clip,keepaspectratio]{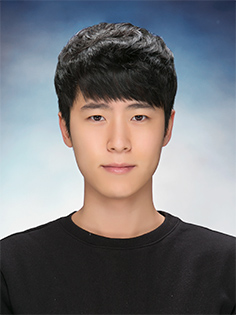}}]{Jaeheum Lee}
received the Bachelor’s degree in Information and Communication engineering from Chungbuk National University, Cheongju, South Korea, in 2018. He is currently working toward the M.S. degree in the Department of information and communication Engineering, Chungbuk National University, Cheongju, South Korea.

His research interests are in the field of Artificial Intelligence and communication circuit design.

\end{IEEEbiography}

\begin{IEEEbiography}[{\includegraphics[width=1in,height=1.25in,clip,keepaspectratio]{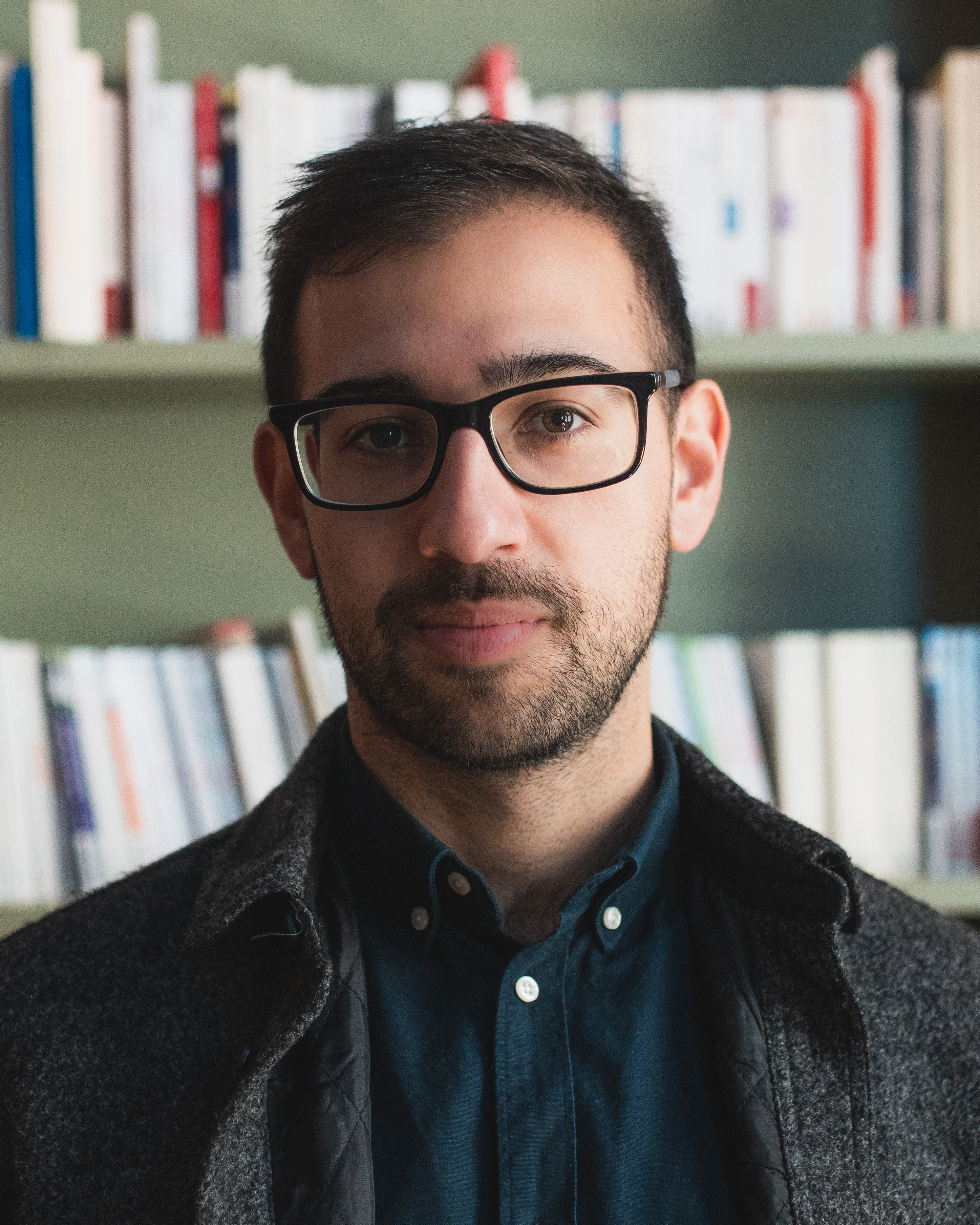}}]{Jason K. Eshraghian}
(S'16-M'19) received the LL.B \& B. Eng, and Ph.D degrees in electrical and electronic engineering from the University of Western Australia, Perth, WA, Australia in 2016 and 2019, respectively.

From 2015 to 2016, he was a Research Associate at Chungbuk National University, Cheongju, South Korea working on the Memristive Retina Project, and joined the University of Michigan, Ann Arbor, as a Post-Doctoral Fellow in 2019. His current research interests include memristive systems, inference acceleration and generative adversarial networks

He was awarded the 2019 IEEE Very Large Scale Integration Circuits and Systems Best Paper Award, and the 2019 IEEE Artificial Intelligence Circuits and Systems Conference Best Paper Award, in recognition for his contributions to the field of neuromorphic computing. He is a member of the IEEE Neural Systems and Applications Technical Committee. 
\end{IEEEbiography}

\begin{IEEEbiography}[{\includegraphics[width=1in,height=1.25in,clip,keepaspectratio]{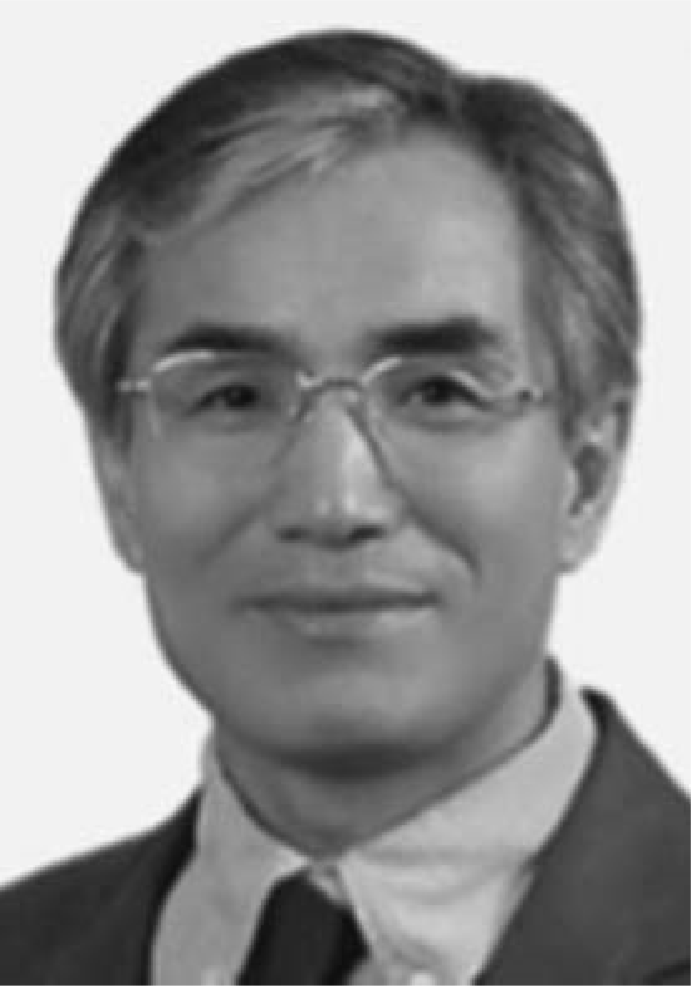}}]{Kyoungrok Cho}
(S'89–M'92) received the B.S. degree in electronic engineering from Kyoungpook National University, Taegu, Korea, in 1977 and the M.S. and Ph.D. degrees in electrical engineering from University of Tokyo, Tokyo, Japan, in 1989 and 1992, respectively. 

From 1979 to 1986, he was with the TV Research Center, LG Electronics, Seoul, South Korea. In 1999 and 2006, he was with Oregon State University, Corvallis, OR, USA, as a Visiting Scholar. He is currently a Professor at the College of Electrical and Computer Engineering, Chungbuk National University, Cheongju, South Korea, where he is also the Director of the IC Design Education Center. His current research interests include highspeed and low-power circuit designs, SoC platform design for communication systems, and prospective CMOS image sensors, memristor-based circuits, and the design of multilayer system-on-systems technology.

He is currently a Professor in the College of Electrical and Computer Engineering, Chungbuk National University, Cheongju, Korea, where he is also a Director of the World Class University program.

\end{IEEEbiography}
\begin{IEEEbiography}[{\includegraphics[width=1in,height=1.25in,clip,keepaspectratio]{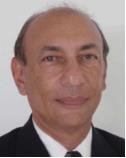}}]{Kamran Eshraghian}
is the Executive Chairman and President of iDataMap Corporation, Australia. He received his Ph.D. degree from University of Adelaide South Australia in 1980 and was also awarded the Doctor of Engineering (Dr.-Ing e.h.) from the University of Ulm, Germany in 2004, for his research into integration of nanoelectronics with that of light wave technology.

In 1994, he was invited to take up the Foundation Chair of Computer, Electronics and Communications Engineering in Western Australia, and became the Dean of School of Engineering and Mathematics and Distinguished University Professor and subsequently became the Director of Electron Science Research Institute. In 2004 he became founder/President of Elabs (Eshraghian Laboratories) as part of his vision for horizontal integration of nanoelectronics with those of bio and photon-based technologies, thus creating a new design domain for System on System (SoS) integration. In 2007 he was the holder of inaugural Ferrero Family Chair in Electrical Engineering and visiting Professor of Engineering at University of California, Merced prior to his move to CBNU, Korea as a Distinguished Professor to lead the World Class University program in the memristor-based SoS arena.  He has co-authored six textbooks and has lectured widely on multi-technology systems. He has over 40 patents and founded six high technology companies, providing intimate link between university research and industry.

His current research interest is on the frontier of computational neuroscience integrated systems. Prof. Eshraghian is a Fellow and life member of the Institution of Engineers, Australia.

\end{IEEEbiography}




\end{document}